\documentclass[12pt,preprint]{aastex}

\slugcomment{appearing in Astron. J., July 2009, 
vol. 138, pp. 240-250}

\shorttitle{Small Trojans with Spitzer}
\shortauthors{Fern\'andez, Jewitt, and Ziffer}

\begin{document}

\title{Albedos of Small Jovian Trojans}

\author{Yanga R. Fern\'andez}
\affil{Department of Physics, University of Central Florida,\\
4000 Central Florida Blvd, Orlando, FL 32816-2385}

\author{David Jewitt}
\affil{Institute for Astronomy, University of Hawaii,\\
2680 Woodlawn Dr, Honolulu, HI 96822}

\and

\author{Julie E. Ziffer}
\affil{Department of Physics, University of Southern
Maine, \\ 96 Falmouth St, Portland, ME 04104-9300}

\begin{abstract}
We present thermal observations of 44 Jovian Trojan asteroids with
diameters $D$ ranging from 5 to 24 km. All objects were observed
at a wavelength of 24 $\mu$m with the Spitzer Space Telescope.
Measurements of the thermal emission and of scattered optical light,
mostly from the University of Hawaii 2.2-meter telescope, together
allow us to constrain the diameter and geometric albedo of each
body.  We find that the median R-band albedo of these small Jovian
Trojans is about 0.12, much higher than that of ``large" Trojans
with $D > 57$ km (0.04). Also the range of albedos among the small
Trojans is wider.  
The small Trojans' higher albedos are also glaringly
different from those of cometary nuclei, which match our sample
Trojans in diameter, however they roughly match the spread of albedos
among (much larger) Centaurs and trans-Neptunian objects.  
We attribute the Trojan albedos to an evolutionary effect: the
small Trojans are more likely to be collisional fragments and so
their surfaces would be younger. A younger surface means less
cumulative exposure to the space environment, which suggests
that their surfaces would not be as dark as those
of the large, primordial Trojans.  In support of this
hypothesis is a statistically significant correlation of higher
albedo with smaller diameter in our sample alone and in a sample
that includes the larger Trojans.  This
correlation of albedo and radius implies that the true size
distribution of small Trojans is shallower than the visible magnitude
distribution alone would suggest, and that there are approximately
half the Trojans with $D>1$ km than previously estimated.

\end{abstract}


\section{Introduction}

Jupiter's Trojan asteroids inhabit two swarms centered on the L4
and L5 Lagrangian points located 5.2 AU from the Sun and from the
planet.  More than 2700 Trojans are known at the time of writing.
Based on optical studies, 
the total population larger than 1 km in radius has been estimated 
by various workers: \citet{jewitt00} estimated
$\sim$1.6$\times10^5$ such objects in the L4 swarm;
\citet{szabo07} estimated $\sim$2.4$\times10^5$ in
both swarms combined;
\citet{yn05} estimated $\sim$2.4$\times10^5$ in the
L4 swarm; and
\citet{nakam08} estimated $\sim$0.63$\times10^5$ in
the L4 swarm and $\sim$0.34$\times10^5$ in the L5. 
The magnitude-derived 
size distribution resembles a broken power
law \citep{jewitt00}, 
and is such that the bulk of the mass (approximately 10$^{-4}$
M$_{\oplus}$, where M$_{\oplus}$ = 6$\times$10$^{24}$ kg is the
mass of the Earth) is contained within the largest objects.  By
number and by mass, the Trojan population is only slightly inferior
to the population of the main-belt asteroids.   However, the
observational attention given to the Trojans so far is miniscule
compared to that lavished on the main-belt objects, and many of the
basic properties of Jupiter's Trojans remain poorly known.  The
Trojans have been reviewed alongside the irregular satellites of
Jupiter, to which they may be closely related, by \citet{jewitt04}
and separately by \citet{dotto08}.

Scientific interest in the Trojans focuses both on their origin and
on their composition.  How and when they were trapped in 1:1
mean-motion resonance with Jupiter remains unknown.  Capture at a
very early epoch in association with planet formation and capture
much later, in a dynamical clearing phase in the Solar system, are
both under current consideration \citep{morby05,marzari07}.  The
snow-line in the Solar system was most likely inside the orbit of
Jupiter \citep{gl07}, so if they formed in-situ or at a more distant
location in the Sun's protoplanetary disk, the Trojans could have
incorporated water as bulk ice.  In this sense, the Trojans might
share as much in common with the nuclei of comets as with the
classical, rocky asteroids.  Observationally, the measured Trojans
resemble the nuclei of short-period comets in their optical colors
\citep{jl90,fornasier07} and albedos \citep[Paper I]{fern03}, tending
to reinforce by association the possibility that they might be
comet-like, ice-rich bodies.  On the other hand, low-resolution
spectral observations in the near infrared have uniformly failed
to reveal absorption bands that could be attributed to water ice
or, indeed, to show any absorption bands at all
\citep{luu94,dumas98,eb03,yang07}.  The low albedos, neutral to
reddish optical colors and featureless near infrared spectra are
compatible with, but not uniquely diagnostic of, irradiated, complex
organics \citep{cruik01}.

The absence of water ice is easily understood as a consequence of
sublimation, even at Jupiter's distance.  For example, dirty
(absorbing) water ice exposed at the sub-solar point on a Trojan
at 5.2 AU sublimates in equilibrium at a rate of $\dot m \sim
8\times10^{-7}$ kg m$^{-2}$ s$^{-1}$, corresponding to recession
of the sublimating surface at speed $\dot m / \rho \sim 2$ cm
yr$^{-1}$, where $\rho \sim$ 10$^3$ kg m$^{-3}$ is the bulk density.
In a few years, any exposed dirty water ice on a Jovian Trojan would
recede into the surface by a depth greater than the diurnal thermal
skin depth (i.e., approximately 5 to 10 cm on a body rotating with
a period $\sim$0.5 day and a thermal diffusivity $\kappa \sim
10^{-7}$ m$^2$ s$^{-1}$).  Clean (i.e. pure) surface ice could
survive much longer, by virtue of its higher albedo and lower
temperature, but sustaining clean surface ice will be difficult in
the face of micrometeorite gardening and contamination.  Just as
with the nuclei of comets, then, the Trojans could have ice-rich
interiors but relatively (or even, completely) ice-free surfaces
composed of refractory, particulate matter (``mantles'').  In this
case, it is possible that collisions within each swarm \citep{marzari02}
occasionally cause previously-embedded and relatively pristine
material to be exposed to space. While no water ice has been
definitively detected spectroscopically on the surfaces of larger
Trojans (as mentioned above), smaller bodies, currently just beyond
the range of ground-based spectroscopic observation, may hold some
remnant near-surface ice.

In order to address these topics, we are investigating some of the
physical properties of the known Trojans. In earlier work (Paper
I), we found that the geometric albedos of Trojans larger than
$\sim$60 km in diameter (``large" Trojans) are uniform.  The
mean R-band geometric albedo of such objects is $0.044\pm0.002$ and
the standard deviation is just 0.008.
(These are transformed from the paper's V-band results using the
average color derived by \citet{fornasier07} of $V-R = 0.45$.)
In our sample of 32 objects,
there was only one outlier (4709 Ennomos), 
with an albedo of 0.14$\pm$0.02.  We
interpreted this uniformity in reflectance to be indicative of
mostly similar evolutionary history across the large end of the
Trojan distribution. Not all such Trojans are exactly the same -- as
shown, for example, in the distribution of visible- and near-infrared
spectral slopes \citep{fornasier07} -- but the narrow spread in albedos
lies in stark contrast to other outer Solar System populations such
as Centaurs and trans-Neptunian objects \citep{stansber08}.
Interestingly, the albedo of large Trojans most closely matches
the cometary nuclei \citep{lamy04} even though there is a large
size mismatch.

In this paper we report results from our program to study the albedos
of ``small" Trojans that more closely match the comets in diameter.
In \S 2 we present our observations, in \S 3 we discuss
the interpretation, and in \S 4 we discuss some implications
of our work.

\section{Observations}

We have two datasets, one obtained with the Spitzer Space Telescope
\citep[SST,][]{werner04} that provided us with mid-infrared imaging,
and another with the University of Hawaii 2.2-meter Telescope that
provided us with visible-wavelength imaging. Table 1 provides a
list of our targets and the circumstances of the observations. The
targets were chosen to have excellent ephemerides so that there
would be no doubt about the success of the SST observations. At the
time we prepared the project, our targets were among the smallest
numbered Trojans known (as judged by $H$, the absolute magnitude).

\subsection{Spitzer Data}

We used the Multiband Imaging Photometry for Spitzer
\citep[MIPS,][]{rieke04} aboard SST to observe all 44 small Trojans
during Cycle 1. Each Trojan was observed in ``photometry" mode using
the 24-$\mu$m imager (effective wavelength $\lambda=23.68\ \mu$m),
a 128-by-128 array of Si:As impurity band conduction detectors.
The scale is 2.55 arcseconds per pixel, and the spatial
resolution is diffraction-limited (Rayleigh criterion of 7.1 arcsec).  The integration time
was 132 seconds,  using 3-second exposure times and 3 cycles,
resulting in 44 individual raw exposures.  Each visit to a Trojan
lasted 6.7 minutes, including observing overheads.  Raw data was
processed by the SST pipeline version 14.4.0 to produce flux-calibrated
``BCD" (basic calibrated data) images. A discussion of the pipeline
processing is given by \citet{gord05}. In general, the
data quality was high and the Trojans provided good
signal-to-noise ratios in the individual frames. No latents
or streaks were seen. 

To measure the flux density of each Trojan, we used two
independent methods. First, we used MOPEX \citep{mm05} to 
obtain photometry
of an object using its individual BCD images. The targets
were bright enough that stacking to boost the signal
was not necessary.
This gave us 44 separate samples of
a Trojan's brightness, with which we could calculate an 
appropriate mean and error, and also readily identify
bad frames. 
Second, we used Interactive Data Language (IDL) software 
to analyze post-BCD mosaics provided by
the SST pipeline. These post-BCD data are combinations
of the BCD images and have had
array distortions rectified \citep{mdh07}. 

In both methods, aperture
photometry was performed usually using an aperture of radius
3.0 BCD pixels (7.65 arcsec), 
though reduced to 2.0 or 2.5 BCD pixels when the 
Trojan was near a background object.  The results
were compared and in all cases the differences
were at the few percent level. Averages and propagated
errors were then calculated.

This photometry  was then corrected for aperture
loss and for color to produce a final measurement of the monochromatic
flux density.  All Trojans appeared as point
sources in all images, facilitating an aperture correction.
Color corrections were calculated from the shape of
the expected spectral energy distribution that results from the
thermal model (see \S 3) and the known Trojan-Spitzer-Sun angles
and distances.

Our photometry is listed in Table 2, with 1$\sigma$ error bars.
Errors in the photometry result from uncertainty in the photon
counting, in measuring an appropriate
sky background, and in the repeatability of the photometry
from BCD to BCD.

\subsection{UH 2.2-meter Telescope Data}

Optical photometry was obtained on the nights of UT 2005 April 7,
April 8, June 28, June 29, and June 30 using the University of
Hawaii 2.2-meter Telescope located atop Mauna Kea, Hawaii.  We used
a Tektronix charge-coupled device (CCD) camera located at the f/10
Cassegrain focus to image the Trojans through an R-band filter
approximating the Kron-Cousins photometric system.  Image scale
with this set-up was 0.219 arcseconds per pixel. The image quality
delivered by the telescope, including the effects of the atmosphere
and wind-shake of the telescope, was typically 0.8 to 1.0 arcseconds
full width at half maximum (FWHM).

Photometric calibration was obtained using observations of standard
stars from the list by \citet{landolt92}, giving us effectively
Cousins R-band magnitudes.  We selected the faintest standard stars
and those having broadband colors most similar to the Sun in order
to minimize photometric uncertainties owing to the shutter and to
color terms introduced by the use of broadband filters.  We also
observed the standards at airmasses similar to the airmasses of the
Trojans, to minimize atmospheric extinction corrections.  The sky
on all nights was photometric except for part of the night of April
7, as judged by the real-time data from the ``Skyprobe'' instrument
on the Canada-France-Hawaii Telescope.  Data taken through thin
clouds on April 7 were calibrated using the photometry of the same
field stars observed on April 8.

Photometry was performed using concentric, circular projected
apertures, typically from 4 to 7 pixels (0.9 to 1.5 arcseconds) in
radius. Several of our targets were observed at low Galactic latitude
and so we took care to select an aperture size and sky location so
as to exclude flux from background stars.  Integration times employed
were short enough that trailing of the Trojans relative to the fixed
stars was comparable to, or less than, the image FWHM, so resulting
in no photometric consequence.

Aperture and color corrections were applied to our photometry and
the resulting final Cousins R-band magnitudes are listed in Table
2, with 1$\sigma$ error bars.  Note that for 12 of our 44 objects,
optical data were not obtained or were unusable due to stellar
crowding.  Error in the photometry results mainly from uncertainty
in the aperture correction and in the determination of an appropriate
sky background.

\section{Physical Parameters}

\subsection{Thermal Model}

The basic radiometric method to obtain an effective diameter $D$
and geometric albedo $p$ is to solve two equations with these two
unknowns, first done many years ago \citep{allen70,matson71,morrison73}
and described in detail by (e.g.) \citet{lebspen89}. One must observe
the reflected sunlight (usually in visible wavelengths) and the
thermal emission (usually in mid-infrared wavelengths); the former
is proportional to $D^2 p$, while the latter is proportional to
$D^2 (1-pq)$, where $q$ is the phase integral.  In our study, we
observed Trojans only in Cousins R-band, so the geometric albedo
is specific to that band and represented by $p_R$.

The method requires knowing the distribution of temperature across
the object's surface, which itself depends on many parameters
including the orientation and magnitude of the spin vector and
the thermal diffusivity/thermal inertia of the surface materials.
The spin vectors and thermal properties of the sample Trojans are
unfortunately unknown.  The median rotation
period for Main-Belt asteroids of the appropriate diameter scale
is about 6 hours \citep{ph04}. Thermal inertias of
primitive asteroids are less well studied, but
recent work
on cometary nuclei and Centaurs suggest
that their thermal inertias are roughly $\sim$10 J/m$^2$/K/$\sqrt{\rm s}$
\citep[e.g.][]{fern06,grou07,li07,lamy08,grou09}.
These parameters, if applicable to small Trojans, indicate that at 5 AU
the Trojans would lie in the ``slow rotator" regime
\citep[cf.][]{spe89}.

The thermal model that we have employed to 
interpret our data is the ``NEA Thermal
Model" (NEATM) devised by \citet{harris98}, 
a simple and widely-used modification to the older
``standard thermal model" \citep[STM;][]{lebspen89}.  
The STM and NEATM generally
apply if the rotation is so slow or the thermal
inertia so low that every point on the surface is 
near instantaneous
equilibrium with the impinging solar radiation. In the case of zero
thermal inertia, the
temperature is a maximum at the subsolar point and decreases as
$(\cos\vartheta)^{1/4}$, where $\vartheta$ is the local solar zenith
angle.

To use NEATM we must make some assumptions.
We assume that emissivity
$\epsilon$ = 0.9 and the phase slope parameter $G$ = 0.05.
We also
assume a value for the beaming parameter, $\eta$, which is a rough
proxy for thermal inertia and the effects of surface roughness,
night-side emission, and beaming from e.g. craters.
In Paper I we found that $\eta=0.94$ was an appropriate average
value for the large Trojans, so we employ that value again here.
We note that recent work \citep[e.g.][]{hd99,delbo03,delbo07}
indicates that small bodies can have a variety of values for $\eta$,
and that the beaming
parameter is often strongly dependent on the phase angle. Fortunately,
all of our sample objects were observed at similar low phase angles.
We address in \S 4 the effect that changing $\eta$ would have on
our results. 

\subsection{Non-simultaneity}

Technically the derivation of diameter and albedo from this method
requires that the observations be done simultaneously, or at least
while knowing the rotational context of the observations.  Neither
condition was satisfied by our datasets, since it is difficult to
schedule ground-based observations to match SST observations. This
means that the diameters and albedos that we derive may not be
exactly correct for a specific object.  Depending on the different
rotational phases at which the thermal and reflected signals are
measured, the derived diameters and albedos could be either too
high or too low by an amount that depends on the deviation of the
shape from spheroidal.

Fortunately, this effect should average out. Our sample size is
large enough, and we have detected all of our targets at significant
signal-to-noise ratio so that we are not missing the faint end of
the sample. Thus, we have an  approximately equal number of Trojans
with both too high and too low albedos. While
the albedos for individual objects may be off from their
true values, the average and
median of the ensemble of apparent albedos should be close to the true
average and median. The spread of the distribution will be wider
than it really is, but the extent of this spread can be estimated
(see \S 4).

In any case, to make use of the multiwavelength photometry we had
to convert the visible magnitudes to account for the differing
heliocentric distance $r$, geo/Spitzercentric distance $\Delta$,
and geo/Spitzercentric phase angle $\alpha$. In other words, we
needed to estimate what each Trojan's magnitude would be had it
been observed by the UH 2.2-meter Telescope at the same geometry
at which it was observed by SST.  The correction to the measured
magnitude is $5\log(r_i/r_v) + 5\log(\Delta_i/\Delta_v) +
\Phi(G,\alpha_i)-\Phi(G,\alpha_v),$ where subscripts ``v" and ``i"
refer to the visible and infrared observations, $\Phi$ is the phase
function, and $G$ is the phase slope parameter.

For the 12 objects with no visible-wavelength data, we have used
the absolute magnitude $H$ (as given by the Minor Planet 
Center\footnote{URL 
{\tt http://cfa-www.harvard.edu/iau/lists/JupiterTrojans.html}})
and the average Trojan $V-R$ color \citep[0.45,][]{fornasier07} to
predict what the visible magnitude would be. We assumed an uncertainty
of $\pm$0.1 mag for $H$.

\subsection{Modeling Results}

Since there are two measurements and two parameters to be fit, there
are no degrees of freedom with which to use, say, a $\chi^2$-statistic.
Therefore we employed a Monte Carlo method with which to estimate
the uncertainties of $D$ and $p_R$ based on the uncertainties in
the photometry. For each pair of photometric points -- one mid-IR
and one visible -- we created 500 pairs of hypothetical measurements
distributed normally about the actual measured values and with
sigmas equal to the actual error bars. We then derived the appropriate
$D$ and $p_R$ that fits each pair, giving us 500 pairs of $D$ and
$p_R$.  The means and standard deviations of these distributions
of $D$ and $p_R$ essentially became our ``best-fit values" and
``error bars."

For the 12 objects with no visible-wavelength data, we effectively
have only the one visible data point derived from $H$. For the other
32 objects, however, there are multiple visible data points. For a
Trojan with $N$ such visible measurements, we paired each measurement
in turn with the single mid-IR measurement to create $N$ estimates
for both $D$ and $p_R$ using the Monte Carlo idea described above.
We then averaged together all the estimates to create a single
overall estimate of diameter and albedo. We also propagated the
errors except in cases where the variance of either $D$ or $p_R$
among the $N$ estimates was significantly larger than the nominal
error estimate, in which case we simply used the standard deviation
itself.

Our final values of diameter and albedo are given in Table 3.  The
table includes all 32 objects with multiwavelength data, as well
as the 12 with only infrared data. It is important to note
however that the table's values
do not account for the non-simultaneity of the IR and visible data,
and that an individual radius and albedo may be off due
to the lack of rotational context. The error bars, likewise, do
not include any such systematic effects. We discuss
this further in the next section.

\section{Discussion}

\subsection{Ensemble Properties and Correlations}

A plot of diameter vs. albedo from Table 3 is shown in
Fig. 1. The most striking feature is the evidence of a trend where
the smaller Trojans have higher albedos, or at least a higher
likelihood of having higher albedos. 
The Spearman rank-order correlation coefficient among the diameters
and albedos of these 44 objects is $-0.493$, which corresponds to
a probability of the two quantities being uncorrelated of only
$6.7\times10^{-4}$.  In terms of the sum-squared difference of the
ranks the correlation is significant at the $3.2\sigma$ level.

Since an absolute magnitude reported by the MPC (or by other
agencies for that matter) could potentially be more uncertain than
the 0.1 mag we have arbitrarily assumed --
owing to uncertainty in color transformations, in phase darkening
laws, and in weighting schemes  -- we have
also analyzed the statistical significance of the apparent
trend in Fig. 1 while excluding the 12 objects for which
we did not have our own visible data. We believe
this is a more robust analysis since it uses the results of more
uniform datasets.  In this case,
the Spearman rank-order correlation coefficient is $-0.610$, which corresponds to
an even lower probability of non-correlation of
$2.1\times10^{-4}$.  The significance of the sum-squared
difference of the ranks is even higher, $3.4\sigma$.
So the trend is statistically significant regardless of whether we include
32 or 44 objects.

In Fig. 2 we add the 32 albedos from our earlier work (Paper I) onto
the same plot. Including these data with
the best 32 gives us 64 data points, and
the correlation appears even more pronounced.  The Spearman
rank-order correlation coefficient among the diameters and albedos
of these 64 objects is $-0.852$, which corresponds to a probability
of the two quantities being uncorrelated of only $4.1\times10^{-19}$.
In terms of the sum-squared difference of the ranks the correlation
is significant at the $6.8\sigma$ level.

It is clear from Fig. 2 that the average albedo of a small Trojan
is larger than that of a large Trojan. This is more readily
demonstrated in Fig.  3, where the histograms of the two populations
are compared.  We can also see that the range of albedos is larger.
Note that our earlier work (Paper I) reported V-band albedos, so
we have scaled those albedos to R-band by using the average Trojan
color $V-R = 0.45$ \citep{fornasier07}.  
To be quantitative, we compare the averages, medians, and
standard deviations of the two populations in Table 4. (We have
listed separately the values for our whole sample of 44 and those
values for the 32 objects that have multiwavelength data.)
Clearly the
typical small-Trojan albedo is higher than that of the large Trojans.

Table 4 and Fig. 3 indicate that
the range of albedos among small Trojans is wider
than the range for the large Trojans, but we must be careful  since the
width of the distribution is at least partly artificial due to the
lack of simultaneity in our datasets as described in \S 3.
We can estimate how much of this spread is real based on a study
of Trojan light curves by \citet{mann07}.  They observed 114
Trojans with sparse sampling and derived a distribution of
photometric ranges. Looking at just their 26 Trojans with apparent
diameters under 35 km (so as to approximately match the diameters
of Trojans in our sample), their distribution has a broad peak with
ranges $\Delta m$ from 0 to 0.3 mag. The distribution then tails
off toward $\Delta m = 0.8$ mag. The average range is $\Delta m = 0.24$
mag and the median is $\Delta m = 0.22$ mag. From this we take
$0.24$ mag to be the appropriate average $\Delta m$ for the Trojans
in our sample.  That means that an optical magnitude would be at
most $\pm0.24$ mag different 
from what it would have been had the observation
been taken simultaneously with the Spitzer observation. The average
offset would be somewhat less, approximately half of this, since
it is unlikely that we would have observed each Trojan at a maximum
in the light curve with one telescope and at a minimum with the
other telescope.  However we leave the offset at $0.24$ mag as
a worst-case scenario, corresponding to a change in visible
flux density by a factor of $10^{0.4\times(\pm0.24)} = 0.80$ or
1.25.  To first order that would also be the factor change in the
albedo. So, if hypothetically all the small Trojans had a true
albedo of exactly 0.100, then we would expect to see a distribution that
ranges from ($0.100\times0.80 =$) 0.080 to ($0.10\times1.25=$)
0.125.  Clearly the histogram of small Trojan albedos is wider than
this.  In fact the observed albedos from 0.04 to 0.12 could only
be explained with an average $\Delta m$ of about 0.6 mag. So unless
the small Trojans have substantially higher typical axial ratios than
were measured by \citet{mann07}, the spread of small Trojan albedos
really is intrinsically wider than that of the large Trojans.

We searched for correlations between albedo and other properties
of the small Trojans. These comparisons are shown in Fig. 4, where
we plotted albedos against three orbital parameters and three
observed parameters.  The only panel suggesting a correlation  is
the inclination, in which higher albedo Trojans are more likely to
have low inclination. However the Spearman rank-order correlation
coefficient among the 32 multiwavelength objects is only $-0.357$,
which corresponds to a probability of the two quantities being
uncorrelated of $0.045$.  In terms of the sum-squared difference
of the ranks the correlation is significant at only the $2.0\sigma$
level. 
Adding in the 32 large-Trojan albedos from Paper I improves
the correlation, but this is likely to be spurious since 
the Trojans in the two surveys do not have overlapping
inclinations.

\subsection{Discovery Bias}

It is important to consider whether the trend in Fig. 2 is a product
of discovery bias. That is, perhaps we are measuring
higher albedos because such small Trojans are more likely to
be discovered; after all, a Trojan of a given diameter with
0.12 albedo will be 1.09 mag brighter than one with
the same diameter but 0.044 albedo. If the high-albedo 
small Trojans are
near the limit of what can be discovered by asteroid
surveys, then 1.09 mag of difference would render
a hypothetical low-albedo subpopulation invisible. 
On the other hand, the
situation is not quite this simple since a Trojan could
have been discovered at another  lunation when it
was brighter. Furthermore, the unknown rotational period and 
axial ratio make predicting when a Trojan can and
cannot be discovered difficult. 

A simple argument does suggest though that at
least the fraction of high-albedo 
small Trojans (however one wants to define ``high")  
is greater than that fraction among the large Trojans.
Only 3\% (1/32) of the large Trojans from Paper I
have albedos above the median albedo we have
measured here, 0.117.   If  only 3\% of
all small Trojans in reality have albedos above this
value, then the asteroid surveys would have to have
missed a vast population of Trojans with $5<D<25$ km,
a population that is about 17 times larger than what
has actually been discovered. This seems unlikely.

We can test this situation more rigorously however
by assuming an overall albedo
distribution to the small Trojans and then determining
what the {\sl measured} albedo distribution would
be for the asteroids that are actually
discovered by the asteroid surveys. To do this we created
a virtual population of  small Trojans and assigned them
absolute magnitudes $H$ such that the ensemble's
distribution of $H$ matched that for the real Trojans
as measured 
by \citet{jewitt00}. They found that for Trojans
below a diameter of about 50 km, the cumulative
magnitude distribution $N$ as a function of absolute
magnitude $H$ is $N(<H)\propto10^{\alpha H}$,
with $\alpha=0.40\pm0.06$.  For our modeling,
we assumed $\alpha=0.40$ precisely. The number
of objects in the simulation was 1.26 million. 

We then created an albedo probability distribution
$P(p)$ to dictate what V-band albedo each virtual object
would be assigned. (We discuss $P$ further
below.) From this we calculated the diameters
$D$ for all virtual
objects using $D = 10^{-0.2H} \times $ 1329 km$/\sqrt{p}$.

Next, we assigned orbits to all virtual objects using
a five-dimensional distribution of 
Trojan semimajor axis ($a$),
orbital eccentricity ($e$), orbital inclination ($i$),
argument of perihelion ($\omega$), and
longitude of ascending node ($\Omega$). This 5-D
distribution was derived empirically by extracting the
orbital elements as compiled by the 
Minor Planet Center\footnote{URL 
{\tt http://cfa-www.harvard.edu/iau/lists/JupiterTrojans.html}}. 
For ease, we let each virtual object's perihelion time $t_P$ 
be randomly chosen between 1992 January 1 and
2004 January 1, i.e., sometime within a 12-year span
(since 12 years is about one Trojan orbital period).
From the orbital elements we
could calculate each object's heliocentric distance,
geocentric distance, and phase angle over this 12-year interval.
In combination with $H$, and assuming a linear
phase law of 0.04 mag/deg, we then calculated each object's
V-band magnitude $m_V$ over this span. 
This range of dates was chosen since it falls within a
period when the Spacewatch survey was surveying the sky
down to $m_V\approx21$
\citep{jm98,lars07}. In fact Spacewatch either discovered outright
or independently found almost all of the Trojans in our Spitzer sample. 
  
We decided that a 
virtual Trojan in our simulation was considered
`discovered' (and therefore available for
inclusion in our Spitzer survey) if it ever became brighter than $m_V=20$
over the course of its orbit. 
This is a conservative choice in limiting magnitude; since
ours is a simplistic model and does not explicitly take
into account the actual sky coverage by the discovery surveys
or the robustness of their ability to detect low signal-to-noise asteroids,
we decided to pick a magnitude limit somewhat brighter than
Spacewatch's nominal limit. 

The result of the simulation is an ensemble of discovered,
virtual objects that is a subset of the whole group of objects.
We then created a plot of diameter vs. albedo ($D$ vs. $p_R$) for
these discovered objects that can be compared to the plot
of real observations in Fig. 1. (The R-band albedo $p_R$ was
calculated from $p$ by multiplying by 1.076 as in \S 4.1.)
To do this comparison statistically, 
we followed the recipe for a two-dimensional Kolmogorov-Smirnov (K-S)
test as described by \citet{press92}.

To keep the model simple, 
we created the functional form of $P(p)$ as follows:

\begin{equation}
P(p) = \left\{ 
\begin{array}{l l}
  0 & \quad \mbox{if $p\le0$,} \\
  C' e^{-(p-p_0)^2/2\sigma_p^2} + C''\Pi(p_l,p_l+p_w;p) & \quad \mbox{if $p>0$,}\\
\end{array} \right. 
\end{equation}

where $\Pi$ is the boxcar function.
In other words, the albedo distribution had a gaussian, low-albedo
component and a uniformly-distributed, high-albedo component.
Specifically, $p_0$ is the mean albedo of
the low-albedo group; $\sigma_p$ is the standard
deviation of the gaussian; $p_l$ is the lower bound of
the high-albedo group; and $p_w$ is the albedo width
of the high-albedo group. 
(Note that for some parameter values, some objects
that belong to the ostensibly low-albedo gaussian could
have albedos that overlap with those from the high-albedo uniform
distribution.)   A fifth parameter,
the fraction of objects with ``high" albedo, $f_h$, controls
the value $C''$:
\begin{equation}
f_h = \int_{p_l}^{p_l+p_w} C''\Pi(p_l,p_l+p_w;p) dp = C'' p_w.
\end{equation}

The overall normalization $\int_{-\infty}^{\infty} P(p) dp = 1$
controls the relative scale of $C'$ and $C''$, so this
setup has five parameters to investigate.
Note that we have not assumed any trend between
diameter and albedo.

Our search through sample space is represented in Fig. 5.
Each panel shows a contour plot of the nominal
probability that our observed plot of $D$ vs. $p_R$
(Fig. 1) and the simulated  plot are drawn from
the same two-dimensional distribution. As \citet{press92}
describe, probabilities greater than about 0.2 may not
be precise due to the simplistic nature of this formulation
of the K-S test, but still do indicate similar distributions.

The similarity of the contours in the panels of Fig. 5
indicates that the `best'
matches are consistently near $p_0\approx0.07$ to $0.12$
and $\sigma_p\approx0.01$ to $0.06$. The
fraction of high-albedo small Trojans $f_h$
in less constrained, but seems
to be roughly under 30\%.
The extent in albedo of that fraction is likewise not well
constrained. This all depends
somewhat on the {\sl a priori} functional form of the distribution
we have assigned, but the important and robust conclusion
is that the small-Trojan albedo distribution  
is definitely not like that of the large Trojans
($p_0\approx0.041$, $\sigma_p\approx0.01$, $f_h\approx3\%$; 
bottom half of Fig. 3);
the probabilities are far too small. This gives us confidence
that the small Trojans really do have different 
surface properties even
accounting for the discovery bias.

\subsection{Effect of Modeling Assumptions}

As stated in \S 3, the phase slope parameter $G$ influences
the final results in Table 3. Re-running our thermal model
for an assumed $G$ of 0.15 instead of 0.05 results in almost
no change to the radii (at the $\sim$10 meter level) and a reduction
in the albedos by about 8 to 9\%. Such a small change would not alter
our conclusions. 

More critical is the choice of $\eta$, since this
certainly can have a significant effect
on the calculated values of both $D$ and $p_R$. To gauge the
influence that our assumptions have on our results, we re-analyzed
our photometry in Table 2 using $\eta=1.25$,
$\eta=1.6$, and $\eta=2.0$
instead of $0.94$. Each value would assume
that the small Trojans had successively higher 
thermal inertia, similar to what has been measured in several
near-Earth asteroids \citep[e.g.,][]{delbo03}. These three trial values
of $\eta$ result in the diameters being (on average) 15\%, 30\%, and 47\% higher
and the albedos being (on average) 24\%, 41\%, 54\% lower than what
we present in Table 3. Thus, if $\eta$ really were 2.0 then the albedos
of the small Trojans would be sufficiently small so that the median
value would approach that of the large Trojans, 0.050 vs. 0.044,
and it would be less clear how significantly more reflective
the small Trojans would be. However, this would mean that
the thermal behavior of the small Trojans would be radically different
from the large Trojans. In effect, an incorrect assumption of
$\eta$ would not nullify the conclusion that the small Trojans
are different from the large ones, it can only alter the
way in which they are different.

But is it likely that smaller Trojans have higher thermal inertia due
to having less regolith, or a large-grained regolith?
The large Trojans ($D>140$ km) seem to have fine-grained silicates on their
surfaces that produce mid-infrared emission bands \citep{emery06},
indicating a fluffy regolith or a regolith where silicates are embedded
in transparent grains. These concepts are also consistent with the average
$\eta$ we found in Paper I. 
There is as yet no such detailed data on small Trojans
(such as those in our current sample) to test whether the regolith 
properties change as a function of size. 

Recent thermal studies of cometary nuclei are suggestive as a
point of comparison, since
the size matching between comets and small Trojans is reasonable,
and since they are both classes of primitive objects. As
mentioned earlier (\S 3.1),  work
by many groups \citep[e.g.][]{jul00,lamy03,grou04,lis05,grou07}
has shown that cometary nuclei have low thermal inertias,
no more than approximately half that of the Moon and often
consistent with zero. Furthermore, work
by \citet{fern08} has shown that the ensemble average
of $\eta$ for
about 50 cometary nuclei observed at 4 to 5 AU from the Sun
is near unity. Such heliocentric distances are very near
that of the Trojans. If the small Trojans are structurally
similar to these comets, then an assumption
of $\eta=0.94$ is quite reasonable.

\subsection{Comparison with Comets}

As noted above, the published properties of the Trojans are broadly
compatible with those of the cometary nuclei.  In particular, the
Trojan optical color distribution resembles that of the cometary
nuclei \citep{jl90} in that both are deficient in ultrared material
known to coat the surfaces of many Kuiper belt objects \citep{jewitt02}.
The albedos of the larger Trojans (Paper I) are likewise similar
to the albedos of cometary nuclei, and suggest a carbonized,
non-volatile surface composition.  Comparison between the physical
properties of the comets and the Trojans is especially interesting
in the context of the Nice model, in which Trojans and Jupiter-family
comets are both products of the Kuiper belt \citep{morby05}.  The
depletion of the ultra-red matter on the comet nuclei and Trojans
already argues either that this supposition is incorrect, or that
the surface properties of Kuiper belt objects are modified after
their removal from the Kuiper belt \citep{jewitt02}.  A systematic
difference in the albedos would demand a similar interpretation.

The new results presented here tend to decrease the similarity with
the comets, in the sense that when Trojans and nuclei \textit{of
the same size} are compared, the Trojan albedos are systematically
higher.  The strength of this statement is limited by the small
sample of cometary nuclei for which reliable albedo determinations
exist, something soon to be corrected by an
on-going survey of cometary nuclei \citep[see][]{fern08}.

\subsection{Size Distribution}

Previous workers derived the size distribution 
of the Trojans based on
the magnitude distribution and
an assumption of constant albedo. \citet{jewitt00}
found a distribution 
consistent with two power laws;  for the largest
objects, with diameters
$D \ge$ 70 km, they found that the differential size distribution's 
power law index is a 
relatively steep $q = 5.5\pm$0.9, but for objects with 
$D$ between 4 and 40 km, they measured $q=3.0\pm0.3$. 
The small-end of the distribution was also measured
by \citet{yn05} and \citet{szabo07}, who found similar
values for $q$: $2.9\pm0.1$ and $3.2\pm0.25$, respectively.
However, now we are in a position to make a better
conversion between absolute magnitude $H$ and 
diameter $D$ since we have a relationship between
$D$ and $p_R$ in Figs. 1 and 2.  The higher albedos found 
for small Trojans imply smaller
diameters than expected, which would
result in a flattening of the 
size distribution relative to the constant
albedo case.

To quantify this effect, we represent the  $D$ vs. $p_R$ trend in Fig. 2 by an 
ad-hoc function. We used the data in Fig. 2 to fit (by least-squares)
the coefficients to the following 4th-order polynomial:
\begin{equation}
p_R(D) = \sum_{m=0}^4 c_m x^m 
\end{equation}
where $x = \log (D/1{\rm\ km})$. However for $D<5$ km
we capped $p_R$ at 0.3, and for $D>143$ km we
set a floor of $p_R = 0.044$. The coefficients are
\begin{equation}
c_0 = 1.540; c_1 = -3.094; c_2 = 2.443; c_3 = -0.858; {\rm\ and\ } c_4 = 0.112.
\end{equation}
This function is plotted with the data in Fig. 2.

We converted the differential size distribution 
provided by \citet{jewitt00} for the L4 swarm
-- their Eqs. 8 and 9 -- back to a luminosity function, i.e. a function of $H$,
using their 0.04 assumed albedo. Using our Eq. 3, we could convert $H$
to a more robust estimate of $D$ and thus then derive a new
size distribution.

The result is shown in Fig. 6. The kink in our size distribution near $D = 5$ km
is due to the break in our $p_R(D)$ function at that diameter. The curvature
to the middle segment between $D=5$ km and $D=35$ km is due to the
curvature in $p_R(D)$ at those diameters, but in log-log space
the segment approximates a power-law.  

The other implication of Fig. 6 is that the number of Trojans larger
than a given size is lower than previously estimated (assuming
that there is no trend of beaming parameter with diameter). Figure 6 indicates
that there 
are approximately $9\times10^4$ Trojans in
the L4 swarm with diameter larger than 2 km,
and about $3\times10^5$ L4 Trojans with diameter larger than 1 km.
This is about a factor of two smaller than the estimate obtained
by \citet{jewitt00} using a constant albedo and $q=3$. 
Other estimates of the Trojan population
\citep[e.g.][]{yn05, szabo07, nakam08} that assume a constant
albedo and use a similar magnitude distribution
would have similar downward corrections to the population
estimate.

\subsection{Origin}

The observed albedo vs. diameter relation could have a number of
causes, ranging from the profound to the insignificant.  The degree
of heating experienced by a solid body due to the decay of embedded
radioactive nuclei increases with the diameter, all else being
equal.  One hundred kilometer scale Trojans will experience a
temperature increase from trapped radio-nuclei approximately 10
times larger than will Trojans only 10 kilometers in scale.  Thus,
it is tempting to think that the observed albedo vs. size relation
might be an artifact of different degrees of metamorphism in the
Trojans, assuming that these objects trapped sufficient quantities
of short-lived radio-nuclei like $^{26}$Al and $^{60}$Fe to be
appreciably heated.  Arguing against this possibility is the size
distribution of the Trojans, which resembles (at least) two power laws
intersecting at about 30 to 40 km diameter (i.e. neatly separating the
sample in the present study from that in Paper I).  Trojans larger
than this  are thought to be survivors of a primordial
population while those smaller than this are more
likely to be products of past, shattering collisions.  If so, the
small objects in the present sample were once part of larger bodies
that must have been radioactively heated, and no simple difference
based on the efficiency with which radiogenic heat can be trapped
is expected.

The albedo vs. diameter relation may instead suggest the action of
some process involving collisions.  The collisional lifetimes of
small Trojans are short compared to the larger objects.  If the
exposed surfaces of Trojan asteroids are progressively darkened,
for example by the irradiation and dehydrogenization of hydrocarbons
 \citep[e.g.][]{thom87,moroz04},
then it is at least qualitatively reasonable to expect an albedo
vs. diameter trend with the sense observed.  

Such a cause might also imply that there
should be a color-diameter
trend in the Trojans, since irradiation can 
change the reflectance slope as well as the albedo. 
Laboratory results indicate that the changes in
the slope depend on dosages and on the original
make-up of the surface \citep[e.g.][]{moroz03,moroz04}, so
there may be no easy answer as to what colors to expect
on Trojans that have suffered various amounts of weathering.
Observationally, \citet{jl90} found
a trend where smaller Trojans (i.e. Trojans with surfaces
that are statistically younger) have redder surfaces, and this
trend was corroborated for D-type asteroids
by \citet{fitz94}, by \citet{lag05}
(for Cybeles), and by \citet{dahl97} (for Hildas). 
Recent work on a wider sample of Trojans has
let some workers study colors of ``background"
Trojans as distinct from those of Trojans
in dynamical families. In particular,
\citet{fornasier07} conclude that there is no
statistically-significant trend between color and
size, while \citet{roig08} argue that among
the background population, it is the larger
Trojans that are redder. In short,
the observational situation regarding a
color vs. diameter relation currently 
remains unresolved.
Future visible
and near-IR datasets on a larger number
of familial and non-familial 
Trojans and on Trojans down to small sizes
may shed more light on this issue.

In any case, in a scenario where collisions play a significant
role in determining the sizes of the small Trojans that now exist in the swarms,
one might expect the size distribution power-law  
to more closely
mimic the Dohnanyi power-law for collisional fragments.
Fig. 6 indicates however
that we have now moved the small-size power-law
to a shallower slope, away from the collisional equilibrium value.
So while collisions likely are influencing the
distribution, there is as yet no simple explanation for Fig. 6.

\section{Summary}

We have measured the 24-$\mu$m 
thermal emission from 44 small Jovian Trojans using the
Spitzer Space Telescope and the R-band reflected
sunlight from 32 of those to derive effective diameters and
albedos. Our sample covers  diameters from 5 to 24 km,
significantly smaller than the large Trojans we sampled in an earlier
survey ($D>57$ km; Paper I). We reach the following conclusions:

\begin{itemize}

\item The measured mean R-band geometric albedo of the small
Trojans in our sample is $0.121\pm0.003$, and the
median is $0.105\pm0.004$. Including only objects for
which we have multiwavelength data, the mean
is $0.137\pm0.004$, and the median is $0.117\pm0.005$. 
These are significantly higher values
than the $0.044\pm0.001$ found for the large Trojans (Paper I).

\item The spread in R-band albedos among the small Trojans
exceeds that of the large Trojans,
with a standard deviation of about $0.065$ vs. $0.008$ (Paper I).

\item The R-band geometric albedo decreases with increasing diameter
in the 5 to 24 km range.  This correlation is significant at the $3.4\sigma$ level, 
and becomes more significant ($6.8\sigma$)
when we include the large Trojans from our earlier work (Paper I).

\item The differences in albedo distribution between the
large and small Trojans are unlikely 
to be caused by either  (a)  the
non-simultaneity in our optical/thermal data, or
(b) a discovery bias toward finding Trojans of high
albedo in the first place. It is possible that 
the albedo differences are artifacts of using
a size-independent infrared beaming parameter
in interpreting the radiometry (and that the
small Trojans have a different ensemble average
thermal inertia than the large ones do), 
but we believe
this possibility to be unlikely. 

\item The origin of the albedo-diameter relation is unknown but
collisions, which shatter and create small bodies on much shorter
timescales than large bodies, may be implicated.

\item The measured size dependence of the albedo tends to flatten the
best-fit power law size distribution index relative to the value computed 
under the assumption of constant albedo.  We find that the differential
power law index that best matches published survey data for objects
in the 5 $\le D \le$ 30 km  range is $q \approx 1.8$, whereas
the value under the constant albedo assumption is $q\approx 3.0$.

\item This flattened size distribution 
implies that there are about a factor of 2
fewer objects of radius greater than 1 km than
estimated when assuming a 0.04 constant albedo.
For example, using the magnitude distribution reported
by \citet{jewitt00}, we find that there
are about $9\times10^4$ L4 Trojans with radius
greater than 1 km instead of the $1.6\times10^5$ inferred
with the constant albedo assumption.

\end{itemize} 

\acknowledgments

We thank John Dvorak for operating the UH telescope, and
for the helpful comments of an anonymous referee and of
Rachel Stevenson.
This work is based in part on observations made with the Spitzer Space 
Telescope, which is operated by the Jet Propulsion Laboratory, 
California Institute of Technology under a contract with NASA. 
Support for this work was provided by NASA through an award 
issued by JPL/Caltech to YRF and DJ, and also through Planetary Astronomy grant
NNG06GG08G to DJ. We acknowledge the reference
material provided by the Minor Planet
Center.

{\it Facilities:} \facility{Spitzer ()}, \facility{UH:2.2m ()}

\clearpage

\begin{figure}
\epsscale{1.00}
\plotone{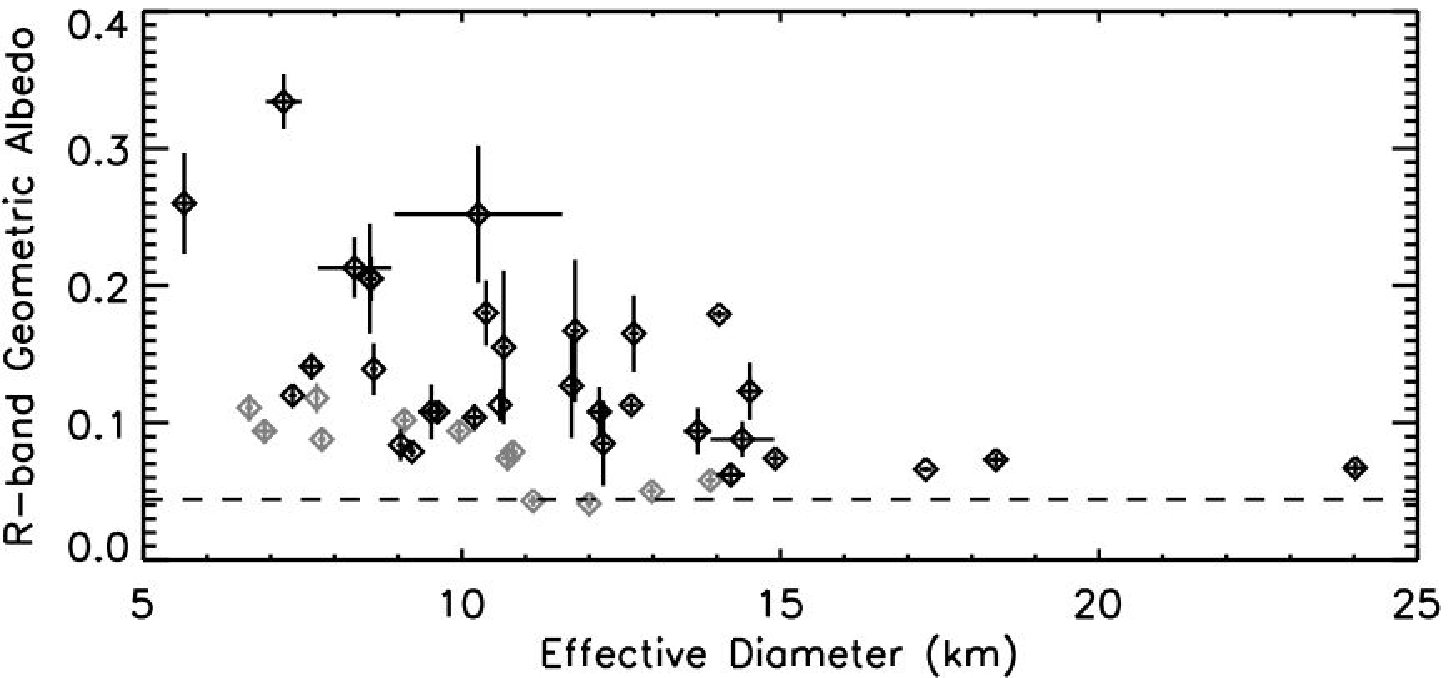}
\caption{Scatter plot of the 44 albedos and diameters derived in
this survey. The 12 objects for which we used $H$ are shown in grey;
the 32 objects for which we have multiwavelength data are shown in
black. The mean albedo of large Trojans as found by us (Paper I),
and translated from V-band to R-band, is indicated with a horizontal
dashed line.  There is a correlation of albedo with radius among
the black points that is significant at the $3.4\sigma$ level.}
\end{figure}

\clearpage

\begin{figure}
\epsscale{1.00}
\plotone{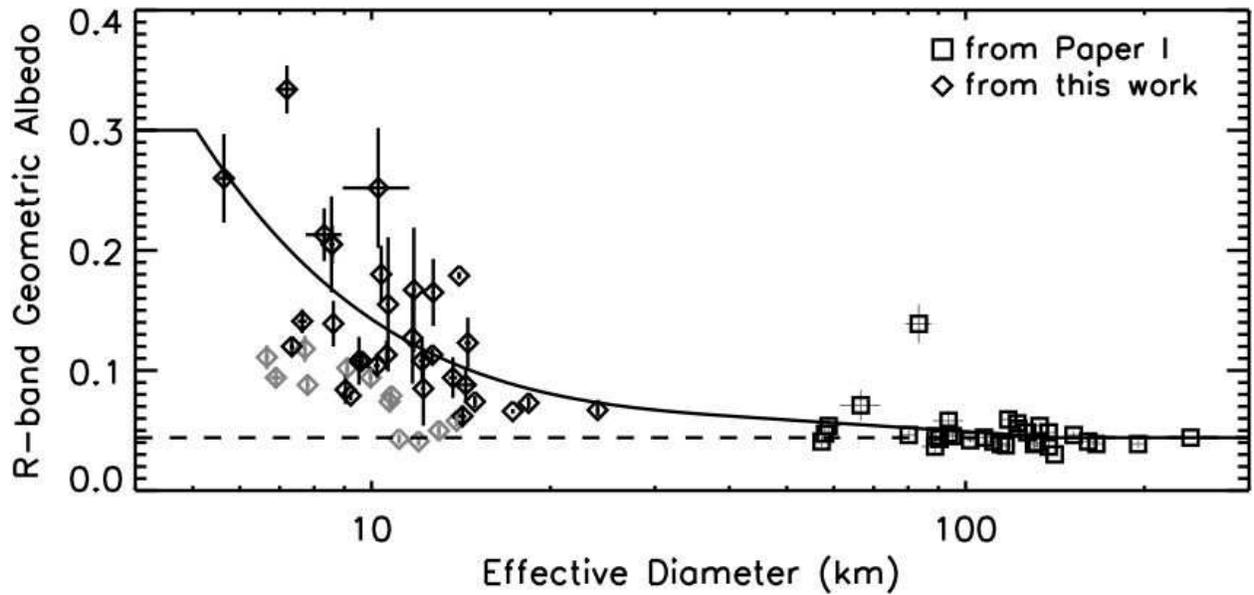}
\caption{Combination of the radii and albedos from the current
survey (diamonds) and from our earlier work (squares; Paper I).
Diamond greyscale is the same as Fig. 1. All 32 points from Paper
I have been included here.  Among the 64 black points there is a
correlation of albedo with radius that is significant at the
$6.8\sigma$ level.  Horizontal dashed line indicates the mean
large-Trojan albedo of 0.044. Solid piece-wise curve represents
an ad hoc function used to investigate the size distribution; see \S 4.5.}
\end{figure}

\clearpage

\begin{figure}
\epsscale{1.00}
\plotone{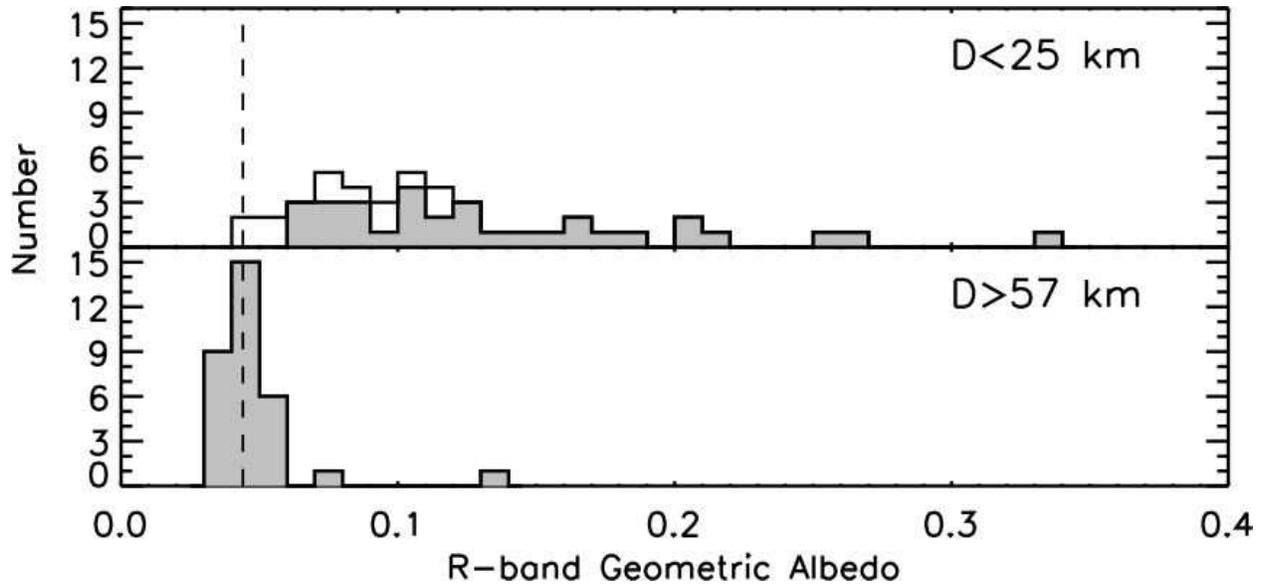}
\caption{Histograms of Trojan albedos. Top panel: distribution of
small Trojan albedos presented in this work. The 32 albedos from
multiwavelength objects are shown with the filled histogram; including
all 44 objects gives the unfilled histogram.  Bottom panel:
distribution of large Trojan albedos reported by us in Paper I. The
means and shapes of the distributions are quite different.}
\end{figure}

\clearpage

\begin{figure}
\epsscale{0.8}
\plotone{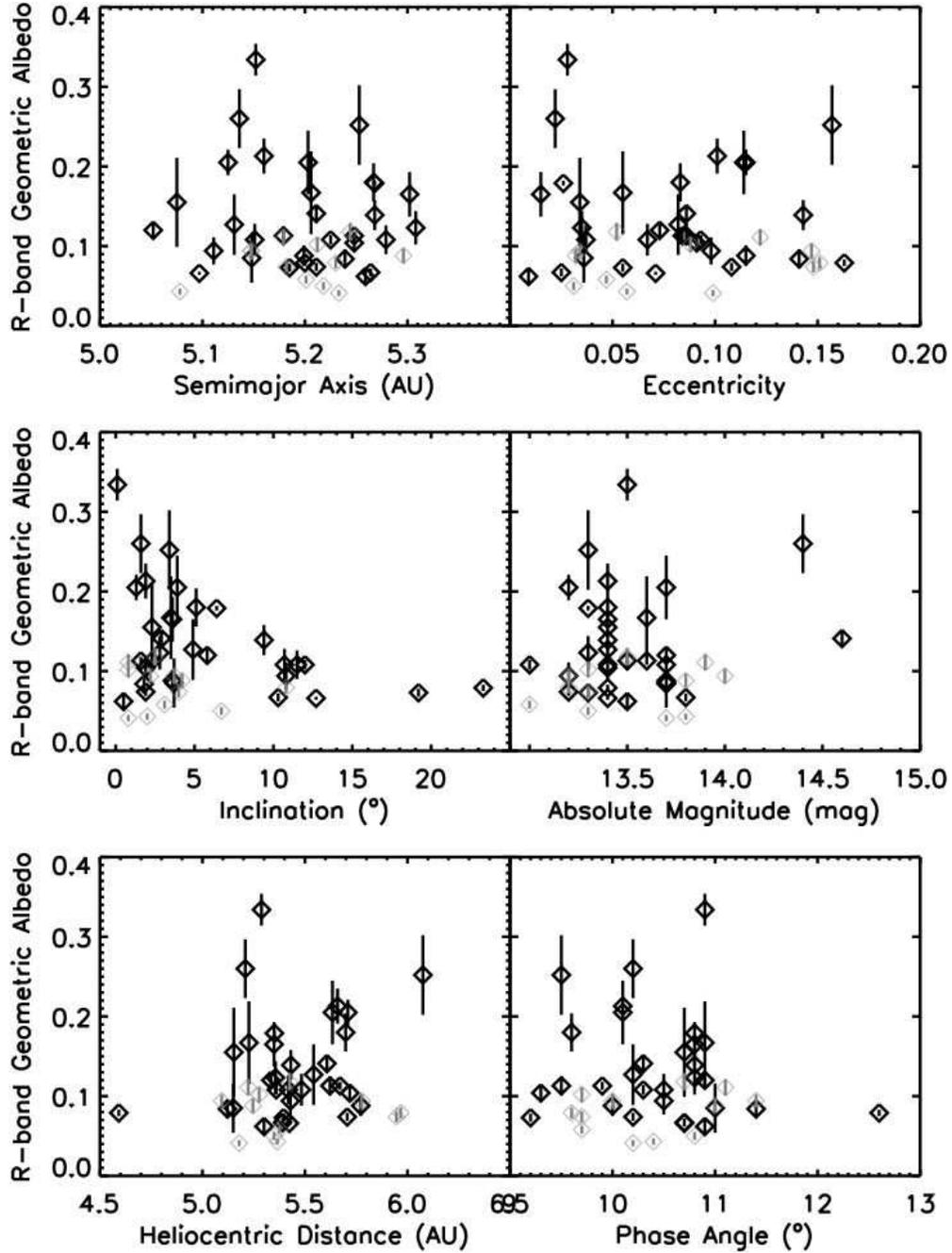}
\caption{Scatter plots of the small Trojan albedos with various
orbital and observed quantities. Diamond greyscale is the same as
Fig. 1. The only panel with some indication of a correlation is the
inclination, but this is significant at only the $2\sigma$ level.} 
\end{figure}

\clearpage

\begin{figure}
\epsscale{0.80}
\plotone{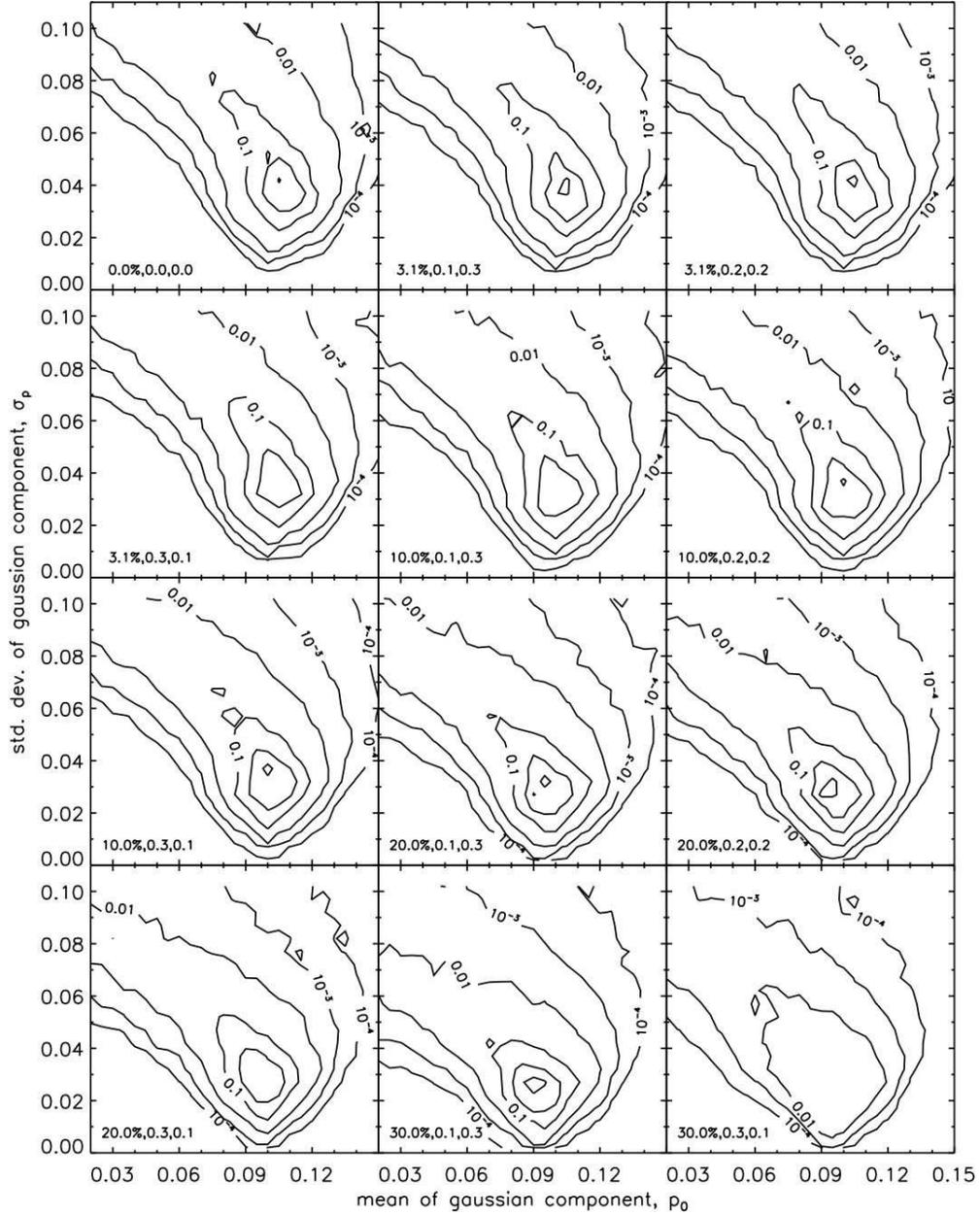}
\caption{Contour plots of the probability that the observed
$D$-vs-$p$ distribution seen in Fig. 1 is drawn from
the same distribution as that based on the simulations
using the five-parameter model described 
in \S 4.2. Each panel represents different values
of $f_h$, $p_l$, and $p_w$; the values are written in
the lower left. Contours correspond to probabilities
of $10^{-4}$ (outermost contour), $10^{-3}$, $0.01$,
$0.1$, $0.3$, and $0.5$.  }
\end{figure}

\clearpage

\begin{figure}
\epsscale{1.0}
\plotone{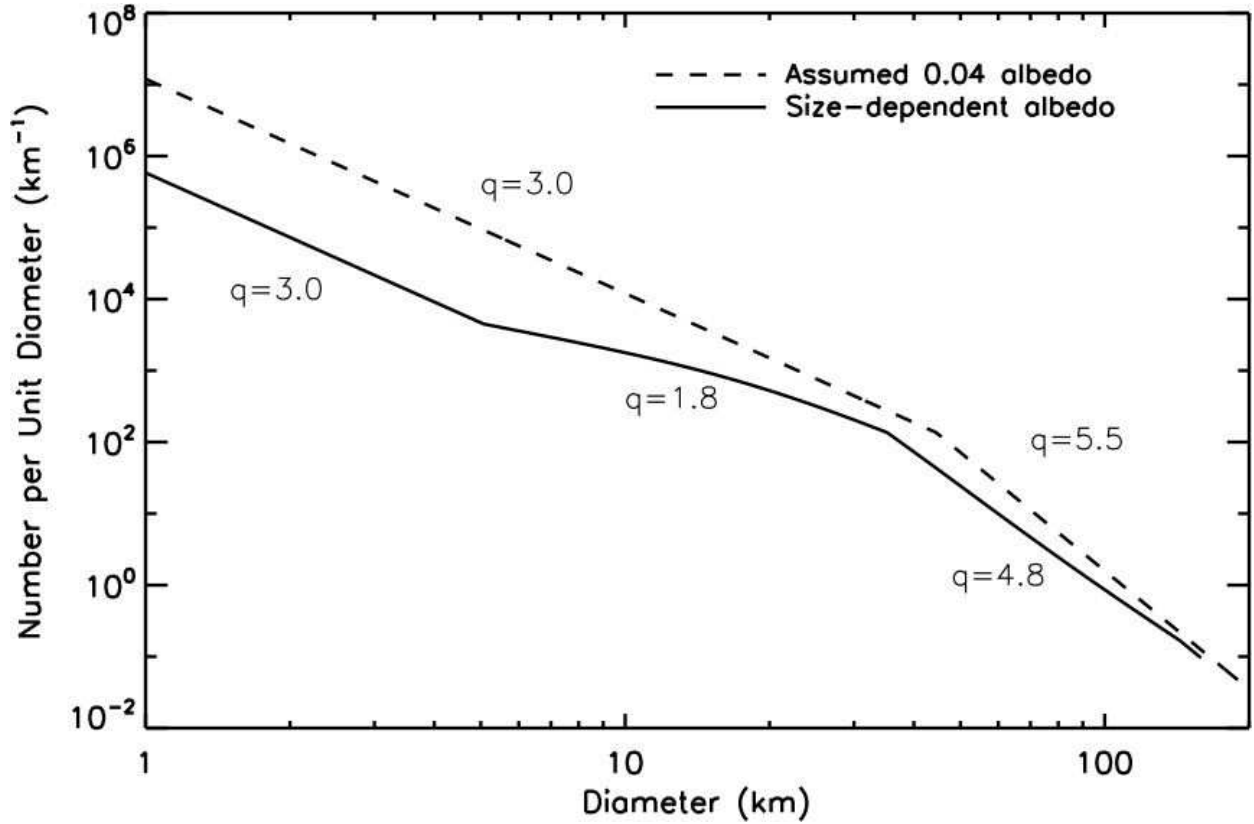}
\caption{Differential size distribution of Trojans. Dashed line is
the distribution derived by \citet{jewitt00} based on an
assumed albedo of 0.04 that was size-independent. Solid line
is our new derivation based on 
their survey data and the size-dependent albedo
shown in Fig. 2. The equivalent power-law slopes, $q$, of each
segment in both distributions are shown.}
\end{figure}

\clearpage

\begin{deluxetable}{cccccllrrr}
\tablecolumns{10}
\tablewidth{0pc}
\tablecaption{Target List and Observing Circumstances}
\tablehead{
\colhead{No.} & \colhead{Name}   & \colhead{L$n$} &
        \colhead{$H$} & \colhead{Tel.} & 
        \colhead{UT Date}   & \colhead{UT} &
        \colhead{$r$ }    & \colhead{$\Delta$ }   & 
        \colhead{$\alpha$} \\
&  &  & \colhead{(mag)} &  & \colhead{yyyy-mm-dd} & \colhead{at start} &
        \colhead{ (AU)}    & \colhead{ (AU)}   & \colhead{ ($^\circ$)}  }
\startdata
 (58153) & 1988 RH$_{11 }$ & L5  & 13.2 &  S &  2004-11-04 & 01:29:25 &  5.706 & 5.506 & 10.2 \\
    "    &      "          & "   &   "  &  H &  2005-04-07 & 07:06:09 &  5.749 & 5.627 & 10.0 \\
    "    &      "          & "   &   "  &  H &  2005-04-08 & 07:44:34 &  5.749 & 5.643 & 10.0 \\
 (37572) & 1989 UC$_{5  }$ & L5  & 13.4 &  S &  2004-11-10 & 08:55:22 &  5.431 & 5.368 & 10.8 \\
    "    &      "          & "   &   "  &  H &  2005-04-07 & 07:11:15 &  5.575 & 5.226 & 10.0 \\
    "    &      "          & "   &   "  &  H &  2005-04-08 & 07:49:39 &  5.576 & 5.242 & 10.0 \\
 (58366) & 1995 OD$_{8  }$ & L4  & 13.7 &  S &  2005-04-08 & 22:17:04 &  5.483 & 5.297 & 10.5 \\
    "    &      "          & "   &   "  &  H &  2005-06-30 & 07:31:15 &  5.473 & 4.499 &  3.4 \\
    "    &      "          & "   &   "  &  H &  2005-06-30 & 09:04:49 &  5.473 & 4.499 &  3.4 \\
 (58475) & 1996 RE$_{11 }$ & L4  & 13.7 &  S &  2005-04-06 & 11:08:11 &  5.150 & 4.835 & 11.0 \\
    "    &      "          & "   &   "  &  H &  2005-04-07 & 12:49:42 &  5.150 & 4.637 & 10.1 \\
    "    &      "          & "   &   "  &  H &  2005-04-08 & 12:11:49 &  5.150 & 4.625 & 10.0 \\
    "    &      "          & "   &   "  &  H &  2005-06-28 & 09:43:02 &  5.173 & 4.219 &  4.3 \\
    "    &      "          & "   &   "  &  H &  2005-06-28 & 10:14:52 &  5.173 & 4.219 &  4.3 \\
(192393)  & 1996 TT$_{22 }$ & L4  & 13.8 &  S &  2005-04-06 & 11:57:26 &  5.246 & 5.229 & 11.0 \\
 (37789) & 1997 UL$_{16 }$ & L4  & 13.5 &  S &  2005-04-08 & 23:20:30 &  5.300 & 5.099 & 10.9 \\
    "    &      "          & "   &   "  &  H &  2005-06-29 & 10:10:33 &  5.303 & 4.306 &  2.4 \\
    "    &      "          & "   &   "  &  H &  2005-06-29 & 12:01:19 &  5.303 & 4.306 &  2.4 \\
\nodata  & 1998 WM$_{24 }$ & L4  & 14.0 &  S &  2005-04-06 & 11:47:36 &  5.777 & 5.729 & 10.0 \\
\nodata  & 1998 WO$_{39 }$ & L4  & 13.4 &  S &  2005-04-06 & 15:51:08 &  5.718 & 5.253 &  9.3 \\
\nodata  &      "          & "   &   "  &  H &  2005-04-07 & 12:32:22 &  5.718 & 5.073 &  8.2 \\
\nodata  &      "          & "   &   "  &  H &  2005-04-08 & 11:18:39 &  5.718 & 5.060 &  8.1 \\
\nodata  &      "          & "   &   "  &  H &  2005-06-28 & 09:18:45 &  5.709 & 4.830 &  5.6 \\
\nodata  &      "          & "   &   "  &  H &  2005-06-28 & 09:52:58 &  5.709 & 4.830 &  5.6 \\
 (40262) & 1999 CF$_{156}$ & L4  & 13.2 &  S &  2005-04-06 & 12:07:00 &  5.967 & 5.981 &  9.6 \\
 (59355) & 1999 CL$_{153}$ & L4  & 13.3 &  S &  2005-05-19 & 13:21:44 &  5.277 & 4.716 &  9.7 \\
 (60257) & 1999 WB$_{25 }$ & L4  & 13.4 &  S &  2005-04-06 & 10:21:48 &  5.153 & 4.763 & 10.7 \\
    "    &      "          & "   &   "  &  H &  2005-04-07 & 12:44:35 &  5.153 & 4.569 &  9.6 \\
    "    &      "          & "   &   "  &  H &  2005-04-08 & 11:57:27 &  5.153 & 4.557 &  9.5 \\
    "    &      "          & "   &   "  &  H &  2005-06-28 & 09:27:00 &  5.171 & 4.250 &  5.2 \\
    "    &      "          & "   &   "  &  H &  2005-06-28 & 10:01:11 &  5.171 & 4.250 &  5.2 \\
 (60322) & 1999 XB$_{257}$ & L4  & 13.8 &  S &  2005-03-10 & 01:30:56 &  5.391 & 5.224 & 10.7 \\
    "    &      "          & "   &   "  &  H &  2005-04-07 & 11:58:09 &  5.391 & 4.655 &  7.8 \\
    "    &      "          & "   &   "  &  H &  2005-04-08 & 10:49:01 &  5.391 & 4.644 &  7.7 \\
    "    &      "          & "   &   "  &  H &  2005-06-28 & 06:41:33 &  5.392 & 4.606 &  7.4 \\
    "    &      "          & "   &   "  &  H &  2005-06-28 & 08:03:45 &  5.392 & 4.606 &  7.4 \\
(192942)  & 2000 AB$_{219}$ & L4  & 13.5 &  S &  2005-04-06 & 11:28:35 &  5.425 & 5.313 & 10.7 \\
 (60388) & 2000 AY$_{217}$ & L4  & 13.8 &  S &  2005-09-23 & 23:13:27 &  5.365 & 4.957 & 10.4 \\
(162396) & 2000 CV$_{120}$ & L4  & 13.0 &  S &  2005-05-13 & 07:11:06 &  5.376 & 4.851 &  9.7 \\
 (60421) & 2000 CZ$_{31 }$ & L4  & 13.3 &  S &  2005-05-19 & 16:00:01 &  5.349 & 5.126 & 10.8 \\
 (62692) & 2000 TE$_{24 }$ & L5  & 13.3 &  S &  2005-04-10 & 07:04:45 &  5.395 & 4.824 &  9.2 \\
    "    &      "          & "   &   "  &  H &  2005-04-07 & 07:25:08 &  5.395 & 4.955 & 10.0 \\
    "    &      "          & "   &   "  &  H &  2005-04-08 & 08:01:59 &  5.395 & 4.969 & 10.0 \\
 (68112) & 2000 YC$_{143}$ & L4  & 13.4 &  S &  2005-04-08 & 03:26:17 &  5.698 & 5.293 &  9.6 \\
    "    &      "          & "   &   "  &  H &  2005-04-07 & 12:39:16 &  5.698 & 5.126 &  8.7 \\
    "    &      "          & "   &   "  &  H &  2005-04-08 & 11:22:54 &  5.698 & 5.112 &  8.6 \\
    "    &      "          & "   &   "  &  H &  2005-06-28 & 09:34:54 &  5.699 & 4.776 &  4.7 \\
    "    &      "          & "   &   "  &  H &  2005-06-28 & 10:08:01 &  5.699 & 4.776 &  4.7 \\
 (63193) & 2000 YY$_{118}$ & L4  & 13.2 &  S &  2005-04-09 & 00:16:01 &  5.429 & 5.182 & 10.5 \\
    "    &      "          & "   &   "  &  H &  2005-04-07 & 13:31:51 &  5.428 & 5.014 & 10.0 \\
    "    &      "          & "   &   "  &  H &  2005-04-08 & 12:57:37 &  5.429 & 5.000 & 10.0 \\
    "    &      "          & "   &   "  &  H &  2005-06-29 & 09:37:46 &  5.471 & 4.490 &  3.1 \\
    "    &      "          & "   &   "  &  H &  2005-06-29 & 12:14:40 &  5.471 & 4.490 &  3.1 \\
 (63259) & 2001 BS$_{81 }$ & L4  & 13.2 &  S &  2005-04-06 & 11:37:43 &  5.094 & 4.979 & 11.4 \\
 (88240) & 2001 CG$_{21 }$ & L4  & 13.4 &  S &  2005-04-08 & 22:06:49 &  5.419 & 5.138 & 10.5 \\
    "    &      "          & "   &   "  &  H &  2005-04-07 & 13:20:37 &  5.418 & 4.972 &  9.9 \\
    "    &      "          & "   &   "  &  H &  2005-04-08 & 12:07:30 &  5.419 & 4.959 &  9.9 \\
    "    &      "          & "   &   "  &  H &  2005-06-29 & 07:57:07 &  5.433 & 4.482 &  4.2 \\
    "    &      "          & "   &   "  &  H &  2005-06-29 & 09:16:37 &  5.433 & 4.482 &  4.2 \\
 (63284) & 2001 DM$_{46 }$ & L4  & 13.3 &  S &  2005-04-08 & 22:57:38 &  6.075 & 5.888 &  9.5 \\
    "    &      "          & "   &   "  &  H &  2005-06-29 & 10:17:41 &  6.067 & 5.071 &  2.1 \\
    "    &      "          & "   &   "  &  H &  2005-06-29 & 12:08:45 &  6.067 & 5.071 &  2.1 \\
 (63279) & 2001 DW$_{34 }$ & L4  & 13.4 &  S &  2005-04-09 & 00:06:01 &  5.659 & 5.401 & 10.1 \\
    "    &      "          & "   &   "  &  H &  2005-06-29 & 08:14:53 &  5.672 & 4.691 &  3.0 \\
    "    &      "          & "   &   "  &  H &  2005-06-29 & 09:26:57 &  5.672 & 4.691 &  3.0 \\
 (28960) & 2001 DZ$_{81 }$ & L4  & 13.3 &  S &  2005-04-08 & 23:39:43 &  5.356 & 5.162 & 10.8 \\
    "    &      "          & "   &   "  &  H &  2005-06-29 & 10:34:59 &  5.336 & 4.338 &  2.3 \\
    "    &      "          & "   &   "  &  H &  2005-06-29 & 11:51:57 &  5.336 & 4.338 &  2.3 \\
(109266) & 2001 QL$_{110}$ & L5  & 13.4 &  S &  2004-12-03 & 14:55:45 &  4.592 & 4.361 & 12.6 \\
    "    &      "          & "   &   "  &  H &  2005-04-07 & 07:35:26 &  4.726 & 4.176 & 10.8 \\
    "    &      "          & "   &   "  &  H &  2005-04-08 & 09:39:42 &  4.727 & 4.192 & 10.9 \\
(156222) & 2001 UB$_{91 }$ & L5  & 13.7 &  S &  2004-11-05 & 18:28:57 &  5.334 & 5.349 & 10.9 \\
    "    &      "          & "   &   "  &  H &  2005-04-07 & 07:18:00 &  5.378 & 5.014 & 10.3 \\
    "    &      "          & "   &   "  &  H &  2005-04-08 & 07:54:47 &  5.378 & 5.030 & 10.4 \\
(156250) & 2001 UM$_{198}$ & L5  & 13.7 &  S &  2004-12-03 & 15:05:49 &  5.122 & 4.933 & 11.4 \\
    "    &      "          & "   &   "  &  H &  2005-04-07 & 07:40:32 &  5.256 & 4.672 &  9.4 \\
    "    &      "          & "   &   "  &  H &  2005-04-08 & 09:43:56 &  5.257 & 4.688 &  9.5 \\
 (64326) & 2001 UX$_{46 }$ & L5  & 13.4 &  S &  2004-12-03 & 14:45:03 &  5.424 & 5.194 & 10.7 \\
    "    &      "          & "   &   "  &  H &  2005-04-07 & 07:47:17 &  5.445 & 4.918 &  9.4 \\
    "    &      "          & "   &   "  &  H &  2005-04-08 & 09:48:22 &  5.445 & 4.932 &  9.5 \\
    "    &      "          & "   &   "  &  H &  2005-06-30 & 06:20:27 &  5.452 & 6.111 &  7.7 \\
    "    &      "          & "   &   "  &  H &  2005-06-30 & 06:26:40 &  5.452 & 6.111 &  7.7 \\
    "    &      "          & "   &   "  &  H &  2005-06-30 & 06:34:27 &  5.452 & 6.111 &  7.7 \\
(158333) & 2001 WW$_{25 }$ & L5  & 13.7 &  S &  2005-05-11 & 09:07:32 &  5.634 & 5.348 & 10.1 \\
    "    &      "          & "   &   "  &  H &  2005-06-30 & 06:05:03 &  5.660 & 6.281 &  7.7 \\
    "    &      "          & "   &   "  &  H &  2005-06-30 & 06:11:14 &  5.660 & 6.281 &  7.7 \\
\nodata  & 2002 CG$_{205}$ & L4  & 14.6 &  S &  2005-03-10 & 16:54:36 &  5.607 & 5.530 & 10.3 \\
\nodata  &      "          & "   &   "  &  H &  2005-04-07 & 12:03:20 &  5.615 & 4.942 &  8.1 \\
\nodata  &      "          & "   &   "  &  H &  2005-04-08 & 11:00:30 &  5.615 & 4.930 &  8.0 \\
\nodata  &      "          & "   &   "  &  H &  2005-06-28 & 07:00:34 &  5.633 & 4.773 &  6.0 \\
\nodata  &      "          & "   &   "  &  H &  2005-06-28 & 08:13:04 &  5.633 & 4.773 &  6.0 \\
 (43627) & 2002 CL$_{224}$ & L4  & 13.2 &  S &  2005-03-10 & 17:55:00 &  5.709 & 5.651 & 10.1 \\
    "    &      "          & "   &   "  &  H &  2005-04-07 & 12:10:05 &  5.710 & 5.052 &  8.1 \\
    "    &      "          & "   &   "  &  H &  2005-04-08 & 11:04:45 &  5.710 & 5.040 &  8.0 \\
    "    &      "          & "   &   "  &  H &  2005-06-28 & 07:08:31 &  5.708 & 4.838 &  5.7 \\
    "    &      "          & "   &   "  &  H &  2005-06-28 & 08:21:08 &  5.708 & 4.838 &  5.7 \\
 (65179) & 2002 CN$_{224}$ & L4  & 13.5 &  S &  2005-04-06 & 15:32:05 &  5.622 & 5.156 &  9.5 \\
    "    &      "          & "   &   "  &  H &  2005-04-07 & 12:15:30 &  5.622 & 4.976 &  8.3 \\
    "    &      "          & "   &   "  &  H &  2005-04-08 & 11:09:01 &  5.622 & 4.964 &  8.2 \\
    "    &      "          & "   &   "  &  H &  2005-06-28 & 07:39:49 &  5.615 & 4.737 &  5.7 \\
    "    &      "          & "   &   "  &  H &  2005-06-28 & 08:49:54 &  5.615 & 4.737 &  5.7 \\
(166115) & 2002 CO$_{208}$ & L4  & 13.9 &  S &  2005-04-10 & 04:40:41 &  5.223 & 5.119 & 11.1 \\
\nodata  & 2002 CS$_{266}$ & L4  & 14.4 &  S &  2005-04-06 & 15:41:46 &  5.209 & 4.734 & 10.2 \\
\nodata  &      "          & "   &   "  &  H &  2005-04-07 & 12:20:35 &  5.209 & 4.555 &  8.9 \\
\nodata  &      "          & "   &   "  &  H &  2005-04-08 & 11:13:16 &  5.209 & 4.542 &  8.8 \\
\nodata  &      "          & "   &   "  &  H &  2005-06-28 & 07:16:28 &  5.219 & 4.342 &  6.2 \\
\nodata  &      "          & "   &   "  &  H &  2005-06-28 & 08:31:21 &  5.219 & 4.342 &  6.2 \\
 (65174) & 2002 CW$_{207}$ & L4  & 13.6 &  S &  2005-04-06 & 10:41:52 &  5.227 & 4.954 & 10.9 \\
    "    &      "          & "   &   "  &  H &  2005-04-07 & 13:15:27 &  5.228 & 4.752 & 10.1 \\
    "    &      "          & "   &   "  &  H &  2005-04-08 & 12:48:53 &  5.228 & 4.738 & 10.0 \\
    "    &      "          & "   &   "  &  H &  2005-06-29 & 07:49:13 &  5.262 & 4.297 &  3.8 \\
    "    &      "          & "   &   "  &  H &  2005-06-29 & 09:08:36 &  5.262 & 4.297 &  3.8 \\
 (65206) & 2002 DB$_{13 }$ & L4  & 13.4 &  S &  2005-04-06 & 10:51:49 &  5.542 & 5.253 & 10.2 \\
    "    &      "          & "   &   "  &  H &  2005-04-07 & 13:01:46 &  5.542 & 5.053 &  9.5 \\
    "    &      "          & "   &   "  &  H &  2005-04-08 & 12:44:37 &  5.542 & 5.039 &  9.4 \\
    "    &      "          & "   &   "  &  H &  2005-06-29 & 07:41:36 &  5.549 & 4.595 &  4.0 \\
    "    &      "          & "   &   "  &  H &  2005-06-29 & 09:00:59 &  5.549 & 4.595 &  4.0 \\
 (89913) & 2002 EC$_{24 }$ & L4  & 13.6 &  S &  2005-04-06 & 10:31:52 &  5.671 & 5.365 &  9.9 \\
    "    &      "          & "   &   "  &  H &  2005-06-29 & 07:34:06 &  5.656 & 4.706 &  4.0 \\
    "    &      "          & "   &   "  &  H &  2005-06-29 & 08:53:29 &  5.656 & 4.706 &  4.0 \\
 (65211) & 2002 EK$_{1  }$ & L4  & 13.5 &  S &  2005-04-08 & 23:29:54 &  5.288 & 5.098 & 10.9 \\
    "    &      "          & "   &   "  &  H &  2005-06-30 & 08:02:27 &  5.282 & 4.285 &  2.4 \\
    "    &      "          & "   &   "  &  H &  2005-06-30 & 09:40:03 &  5.282 & 4.285 &  2.4 \\
(195258)  & 2002 EN$_{52 }$ & L4  & 13.7 &  S &  2005-05-13 & 07:01:27 &  5.179 & 4.678 & 10.2 \\
 (65227) & 2002 ES$_{46 }$ & L4  & 13.3 &  S &  2005-04-12 & 12:34:12 &  5.351 & 5.225 & 10.8 \\
    "    &      "          & "   &   "  &  H &  2005-06-30 & 08:18:12 &  5.363 & 4.355 &  1.6 \\
    "    &      "          & "   &   "  &  H &  2005-06-30 & 09:56:42 &  5.363 & 4.355 &  1.6 \\
 (65217) & 2002 EY$_{16 }$ & L4  & 13.4 &  S &  2005-04-10 & 18:21:17 &  5.349 & 5.232 & 10.8 \\
    "    &      "          & "   &   "  &  H &  2005-04-08 & 13:46:22 &  5.349 & 5.071 & 10.6 \\
    "    &      "          & "   &   "  &  H &  2005-06-30 & 08:26:52 &  5.356 & 4.346 &  1.4 \\
    "    &      "          & "   &   "  &  H &  2005-06-30 & 10:10:29 &  5.356 & 4.346 &  1.4 \\
 (65250) & 2002 FT$_{14 }$ & L4  & 13.3 &  S &  2005-04-13 & 02:29:44 &  5.946 & 5.820 &  9.7 \\
(183358) & 2002 VM$_{131}$ & L5  & 13.0 &  S &  2004-12-03 & 17:38:59 &  5.360 & 4.947 & 10.3 \\
\nodata  &      "          & "   &   "  &  H &  2005-04-07 & 07:30:20 &  5.438 & 5.076 & 10.2 \\
\nodata  &      "          & "   &   "  &  H &  2005-04-08 & 09:34:36 &  5.439 & 5.092 & 10.2 \\
 (58096) & Oineus          & L4  & 13.7 &  S &  2005-04-06 & 11:18:34 &  5.772 & 5.584 & 10.0 \\
    "    &      "          & "   &   "  &  H &  2005-04-07 & 13:40:30 &  5.772 & 5.377 &  9.4 \\
    "    &      "          & "   &   "  &  H &  2005-04-08 & 13:02:50 &  5.771 & 5.362 &  9.4 \\
    "    &      "          & "   &   "  &  H &  2005-06-29 & 10:03:33 &  5.752 & 4.764 &  2.7 \\
    "    &      "          & "   &   "  &  H &  2005-06-29 & 11:42:13 &  5.752 & 4.764 &  2.7 \\
\enddata                     
\tablecomments{Here L$n$ indicates the swarm (Lagrange point) in which each object
resides; $H$ gives the absolute magnitude as listed by the Minor Planet Center
at URL {\tt http://cfa-www.harvard.edu/iau/lists/JupiterTrojans.html};
the ``Tel." column lists which telescope was used: ``S" = Spitzer Space
Telescope, ``H" = University of Hawaii 2.2-meter Telescope;
$r$, $\Delta$, and $\alpha$ list the heliocentric distance,
geocentric/Spitzercentric distance, and
geocentric/Spitzercentric phase angle  as given
by JPL's Horizons system.}       
\end{deluxetable}

\clearpage

\begin{deluxetable}{cccllc}
\tablecolumns{6}
\tablewidth{0pc}
\tablecaption{Photometry}
\tablehead{
\colhead{No.} & \colhead{Name}   & \colhead{Tel.} & 
        \colhead{UT Date}   & \colhead{UT} & \colhead{$F$ or $m_R$ } \\
&  &  & \colhead{yyyy-mm-dd} & \colhead{at start} & \colhead{(mJy or mag)}  }
\startdata
 (58153) & 1988 RH$_{11 }$ & S &  2004-11-04 & 01:29:25 &   $13.54\pm 0.16$ \\
    "    &      "          & H &  2005-04-07 & 07:06:09 &   $20.472\pm0.075$ \\
    "    &      "          & H &  2005-04-08 & 07:44:34 &   $20.567\pm0.072$ \\
 (37572) & 1989 UC$_{5  }$ & S &  2004-11-10 & 08:55:22 &   $ 5.14\pm 0.06$ \\
    "    &      "          & H &  2005-04-07 & 07:11:15 &   $20.725\pm0.074$ \\
    "    &      "          & H &  2005-04-08 & 07:49:39 &   $20.955\pm0.079$ \\
 (58366) & 1995 OD$_{8  }$ & S &  2005-04-08 & 22:17:04 &   $ 6.37\pm 0.28$ \\
    "    &      "          & H &  2005-06-30 & 07:31:15 &   $20.475\pm0.023$ \\
    "    &      "          & H &  2005-06-30 & 09:04:49 &   $20.788\pm0.298$ \\
 (58475) & 1996 RE$_{11 }$ & S &  2005-04-06 & 11:08:11 &   $14.27\pm 0.24$ \\
    "    &      "          & H &  2005-04-07 & 12:49:42 &   $20.523\pm0.126$ \\
    "    &      "          & H &  2005-04-08 & 12:11:49 &   $20.662\pm0.115$ \\
    "    &      "          & H &  2005-06-28 & 09:43:02 &   $19.810\pm0.053$ \\
    "    &      "          & H &  2005-06-28 & 10:14:52 &   $19.767\pm0.034$ \\
(192393)  & 1996 TT$_{22 }$ & S &  2005-04-06 & 11:57:26 &   $ 4.80\pm 0.06$ \\
 (37789) & 1997 UL$_{16 }$ & S &  2005-04-08 & 23:20:30 &   $16.59\pm 0.48$ \\
    "    &      "          & H &  2005-06-29 & 10:10:33 &   $20.211\pm0.254$ \\
    "    &      "          & H &  2005-06-29 & 12:01:19 &   $20.240\pm0.145$ \\
\nodata  & 1998 WM$_{24 }$ & S &  2005-04-06 & 11:47:36 &   $ 2.60\pm 0.13$ \\
\nodata  & 1998 WO$_{39 }$ & S &  2005-04-06 & 15:51:08 &   $ 6.87\pm 0.15$ \\
\nodata  &      "          & H &  2005-04-07 & 12:32:22 &   $20.578\pm0.102$ \\
\nodata  &      "          & H &  2005-04-08 & 11:18:39 &   $20.760\pm0.087$ \\
\nodata  &      "          & H &  2005-06-28 & 09:18:45 &   $20.788\pm0.024$ \\
\nodata  &      "          & H &  2005-06-28 & 09:52:58 &   $20.610\pm0.053$ \\
 (40262) & 1999 CF$_{156}$ & S &  2005-04-06 & 12:07:00 &   $ 5.50\pm 0.08$ \\
 (59355) & 1999 CL$_{153}$ & S &  2005-05-19 & 13:21:44 &   $ 7.94\pm 0.12$ \\
 (60257) & 1999 WB$_{25 }$ & S &  2005-04-06 & 10:21:48 &   $10.96\pm 0.14$ \\
    "    &      "          & H &  2005-04-07 & 12:44:35 &   $20.269\pm0.087$ \\
    "    &      "          & H &  2005-04-08 & 11:57:27 &   $19.316\pm0.087$ \\
    "    &      "          & H &  2005-06-28 & 09:27:00 &   $19.818\pm0.016$ \\
    "    &      "          & H &  2005-06-28 & 10:01:11 &   $19.800\pm0.016$ \\
 (60322) & 1999 XB$_{257}$ & S &  2005-03-10 & 01:30:56 &   $43.57\pm 0.49$ \\
    "    &      "          & H &  2005-04-07 & 11:58:09 &   $19.006\pm0.071$ \\
    "    &      "          & H &  2005-04-08 & 10:49:01 &   $19.107\pm0.074$ \\
    "    &      "          & H &  2005-06-28 & 06:41:33 &   $19.154\pm0.012$ \\
    "    &      "          & H &  2005-06-28 & 08:03:45 &   $19.137\pm0.016$ \\
(192942)  & 2000 AB$_{219}$ & S &  2005-04-06 & 11:28:35 &   $ 4.23\pm 0.05$ \\
 (60388) & 2000 AY$_{217}$ & S &  2005-09-23 & 23:13:27 &   $10.55\pm 0.14$ \\
(162396) & 2000 CV$_{120}$ & S &  2005-05-13 & 07:11:06 &   $17.11\pm 0.20$ \\
 (60421) & 2000 CZ$_{31 }$ & S &  2005-05-19 & 16:00:01 &   $13.46\pm 0.19$ \\
 (62692) & 2000 TE$_{24 }$ & S &  2005-04-10 & 07:04:45 &   $29.93\pm 0.38$ \\
    "    &      "          & H &  2005-04-07 & 07:25:08 &   $19.573\pm0.070$ \\
    "    &      "          & H &  2005-04-08 & 08:01:59 &   $19.594\pm0.066$ \\
 (68112) & 2000 YC$_{143}$ & S &  2005-04-08 & 03:26:17 &   $ 6.90\pm 0.11$ \\
    "    &      "          & H &  2005-04-07 & 12:39:16 &   $20.285\pm0.103$ \\
    "    &      "          & H &  2005-04-08 & 11:22:54 &   $20.183\pm0.109$ \\
    "    &      "          & H &  2005-06-28 & 09:34:54 &   $19.923\pm0.020$ \\
    "    &      "          & H &  2005-06-28 & 10:08:01 &   $19.885\pm0.022$ \\
 (63193) & 2000 YY$_{118}$ & S &  2005-04-09 & 00:16:01 &   $14.13\pm 0.44$ \\
    "    &      "          & H &  2005-04-07 & 13:31:51 &   $20.327\pm0.097$ \\
    "    &      "          & H &  2005-04-08 & 12:57:37 &   $17.747\pm0.076$ \\
    "    &      "          & H &  2005-06-29 & 09:37:46 &   $19.875\pm0.066$ \\
    "    &      "          & H &  2005-06-29 & 12:14:40 &   $19.840\pm0.030$ \\
 (63259) & 2001 BS$_{81 }$ & S &  2005-04-06 & 11:37:43 &   $ 9.11\pm 0.15$ \\
 (88240) & 2001 CG$_{21 }$ & S &  2005-04-08 & 22:06:49 &   $11.30\pm 0.28$ \\
    "    &      "          & H &  2005-04-07 & 13:20:37 &   $20.169\pm0.120$ \\
    "    &      "          & H &  2005-04-08 & 12:07:30 &   $20.466\pm0.122$ \\
    "    &      "          & H &  2005-06-29 & 07:57:07 &   $19.945\pm0.022$ \\
    "    &      "          & H &  2005-06-29 & 09:16:37 &   $19.914\pm0.028$ \\
 (63284) & 2001 DM$_{46 }$ & S &  2005-04-08 & 22:57:38 &   $ 4.66\pm 1.40$ \\
    "    &      "          & H &  2005-06-29 & 10:17:41 &   $20.120\pm0.318$ \\
    "    &      "          & H &  2005-06-29 & 12:08:45 &   $19.908\pm0.097$ \\
 (63279) & 2001 DW$_{34 }$ & S &  2005-04-09 & 00:06:01 &   $ 4.29\pm 0.64$ \\
    "    &      "          & H &  2005-06-29 & 08:14:53 &   $20.409\pm0.023$ \\
    "    &      "          & H &  2005-06-29 & 09:26:57 &   $20.264\pm0.046$ \\
 (28960) & 2001 DZ$_{81 }$ & S &  2005-04-08 & 23:39:43 &   $16.22\pm 0.38$ \\
    "    &      "          & H &  2005-06-29 & 10:34:59 &   $19.308\pm0.241$ \\
    "    &      "          & H &  2005-06-29 & 11:51:57 &   $19.590\pm0.116$ \\
(109266) & 2001 QL$_{110}$ & S &  2004-12-03 & 14:55:45 &   $12.31\pm 0.14$ \\
    "    &      "          & H &  2005-04-07 & 07:35:26 &   $20.577\pm0.079$ \\
    "    &      "          & H &  2005-04-08 & 09:39:42 &   $20.535\pm0.079$ \\
(156222) & 2001 UB$_{91 }$ & S &  2004-11-05 & 18:28:57 &   $ 3.89\pm 0.05$ \\
    "    &      "          & H &  2005-04-07 & 07:18:00 &   $21.173\pm0.074$ \\
    "    &      "          & H &  2005-04-08 & 07:54:47 &   $21.178\pm0.121$ \\
(156250) & 2001 UM$_{198}$ & S &  2004-12-03 & 15:05:49 &   $ 7.58\pm 0.08$ \\
    "    &      "          & H &  2005-04-07 & 07:40:32 &   $21.064\pm0.083$ \\
    "    &      "          & H &  2005-04-08 & 09:43:56 &   $20.844\pm0.083$ \\
 (64326) & 2001 UX$_{46 }$ & S &  2004-12-03 & 14:45:03 &   $22.52\pm 0.26$ \\
    "    &      "          & H &  2005-04-07 & 07:47:17 &   $19.978\pm0.069$ \\
    "    &      "          & H &  2005-04-08 & 09:48:22 &   $19.964\pm0.072$ \\
    "    &      "          & H &  2005-06-30 & 06:20:27 &   $20.470\pm0.042$ \\
    "    &      "          & H &  2005-06-30 & 06:26:40 &   $20.420\pm0.040$ \\
    "    &      "          & H &  2005-06-30 & 06:34:27 &   $20.434\pm0.055$ \\
(158333) & 2001 WW$_{25 }$ & S &  2005-05-11 & 09:07:32 &   $ 4.66\pm 0.07$ \\
    "    &      "          & H &  2005-06-30 & 06:05:03 &   $20.954\pm0.263$ \\
    "    &      "          & H &  2005-06-30 & 06:11:14 &   $20.800\pm0.139$ \\
\nodata  & 2002 CG$_{205}$ & S &  2005-03-10 & 16:54:36 &   $ 3.58\pm 0.13$ \\
\nodata  &      "          & H &  2005-04-07 & 12:03:20 &   $21.010\pm0.097$ \\
\nodata  &      "          & H &  2005-04-08 & 11:00:30 &   $20.968\pm0.097$ \\
\nodata  &      "          & H &  2005-06-28 & 07:00:34 &   $20.906\pm0.026$ \\
\nodata  &      "          & H &  2005-06-28 & 08:13:04 &   $20.974\pm0.025$ \\
 (43627) & 2002 CL$_{224}$ & S &  2005-03-10 & 17:55:00 &   $ 4.08\pm 0.08$ \\
    "    &      "          & H &  2005-04-07 & 12:10:05 &   $20.532\pm0.086$ \\
    "    &      "          & H &  2005-04-08 & 11:04:45 &   $20.381\pm0.089$ \\
    "    &      "          & H &  2005-06-28 & 07:08:31 &   $20.303\pm0.068$ \\
    "    &      "          & H &  2005-06-28 & 08:21:08 &   $20.279\pm0.020$ \\
 (65179) & 2002 CN$_{224}$ & S &  2005-04-06 & 15:32:05 &   $ 7.96\pm 0.22$ \\
    "    &      "          & H &  2005-04-07 & 12:15:30 &   $20.479\pm0.115$ \\
    "    &      "          & H &  2005-04-08 & 11:09:01 &   $20.320\pm0.088$ \\
    "    &      "          & H &  2005-06-28 & 07:39:49 &   $20.439\pm0.019$ \\
    "    &      "          & H &  2005-06-28 & 08:49:54 &   $20.544\pm0.034$ \\
(166115) & 2002 CO$_{208}$ & S &  2005-04-10 & 04:40:41 &   $ 3.65\pm 0.06$ \\
\nodata  & 2002 CS$_{266}$ & S &  2005-04-06 & 15:41:46 &   $ 2.97\pm 0.15$ \\
\nodata  &      "          & H &  2005-04-07 & 12:20:35 &   $20.869\pm0.113$ \\
\nodata  &      "          & H &  2005-04-08 & 11:13:16 &   $20.602\pm0.085$ \\
\nodata  &      "          & H &  2005-06-28 & 07:16:28 &   $20.531\pm0.022$ \\
\nodata  &      "          & H &  2005-06-28 & 08:31:21 &   $20.391\pm0.023$ \\
 (65174) & 2002 CW$_{207}$ & S &  2005-04-06 & 10:41:52 &   $12.03\pm 0.15$ \\
    "    &      "          & H &  2005-04-07 & 13:15:27 &   $19.740\pm0.141$ \\
    "    &      "          & H &  2005-04-08 & 12:48:53 &   $19.169\pm0.123$ \\
    "    &      "          & H &  2005-06-29 & 07:49:13 &   $19.716\pm0.087$ \\
    "    &      "          & H &  2005-06-29 & 09:08:36 &   $19.759\pm0.036$ \\
 (65206) & 2002 DB$_{13 }$ & S &  2005-04-06 & 10:51:49 &   $ 9.59\pm 0.16$ \\
    "    &      "          & H &  2005-04-07 & 13:01:46 &   $20.166\pm0.135$ \\
    "    &      "          & H &  2005-04-08 & 12:44:37 &   $20.713\pm0.116$ \\
    "    &      "          & H &  2005-06-29 & 07:41:36 &   $19.918\pm0.029$ \\
    "    &      "          & H &  2005-06-29 & 09:00:59 &   $19.741\pm0.109$ \\
 (89913) & 2002 EC$_{24 }$ & S &  2005-04-06 & 10:31:52 &   $10.29\pm 0.16$ \\
    "    &      "          & H &  2005-06-29 & 07:34:06 &   $20.110\pm0.022$ \\
    "    &      "          & H &  2005-06-29 & 08:53:29 &   $20.041\pm0.036$ \\
 (65211) & 2002 EK$_{1  }$ & S &  2005-04-08 & 23:29:54 &   $ 3.94\pm 0.35$ \\
    "    &      "          & H &  2005-06-30 & 08:02:27 &   $19.870\pm0.074$ \\
    "    &      "          & H &  2005-06-30 & 09:40:03 &   $19.685\pm0.078$ \\
(195258)  & 2002 EN$_{52 }$ & S &  2005-05-13 & 07:01:27 &   $14.79\pm 0.17$ \\
 (65227) & 2002 ES$_{46 }$ & S &  2005-04-12 & 12:34:12 &   $14.60\pm 0.20$ \\
    "    &      "          & H &  2005-06-30 & 08:18:12 &   $19.098\pm0.018$ \\
    "    &      "          & H &  2005-06-30 & 09:56:42 &   $19.119\pm0.016$ \\
 (65217) & 2002 EY$_{16 }$ & S &  2005-04-10 & 18:21:17 &   $11.97\pm 0.13$ \\
    "    &      "          & H &  2005-04-08 & 13:46:22 &   $19.860\pm0.178$ \\
    "    &      "          & H &  2005-06-30 & 08:26:52 &   $19.411\pm0.018$ \\
    "    &      "          & H &  2005-06-30 & 10:10:29 &   $19.234\pm0.102$ \\
 (65250) & 2002 FT$_{14 }$ & S &  2005-04-13 & 02:29:44 &   $ 5.78\pm 0.10$ \\
(183358) & 2002 VM$_{131}$ & S &  2004-12-03 & 17:38:59 &   $ 7.81\pm 0.09$ \\
\nodata  &      "          & H &  2005-04-07 & 07:30:20 &   $20.704\pm0.081$ \\
\nodata  &      "          & H &  2005-04-08 & 09:34:36 &   $20.705\pm0.081$ \\
 (58096) & Oineus          & S &  2005-04-06 & 11:18:34 &   $11.94\pm 0.79$ \\
    "    &      "          & H &  2005-04-07 & 13:40:30 &   $20.494\pm0.098$ \\
    "    &      "          & H &  2005-04-08 & 13:02:50 &   $20.401\pm0.104$ \\
    "    &      "          & H &  2005-06-29 & 10:03:33 &   $20.056\pm0.055$ \\
    "    &      "          & H &  2005-06-29 & 11:42:13 &   $19.981\pm0.037$ \\
\enddata
\tablecomments{Here 
the ``Tel." column lists which telescope was used: ``S" = Spitzer Space
Telescope, ``H" = University of Hawaii 2.2-meter Telescope; the ``$F$ or $m_R$"
column lists either the flux density $F$ at a wavelength of
23.68 $\mu$m as observed
by Spitzer or the Cousins R magnitude $m_R$ as observed by the UH 2.2-meter Telescope.}
\end{deluxetable}

\clearpage

\begin{deluxetable}{rlcc}
\tablecolumns{7}
\tablewidth{0pc}
\tablecaption{Physical Parameters and Formal Errors}
\tablehead{
\colhead{No.} & \colhead{Name}   & \colhead{$D$ (km)} & 
        \colhead{$p_R$}  }
\startdata
 (58153) & 1988 RH$_{11 }$ &  $14.92\pm0.10$  &  $ 0.074\pm0.005 $ \\
 (37572) & 1989 UC$_{5  }$ &  $ 8.62\pm0.06$  &  $ 0.139\pm0.019 $ \\
 (58366) & 1995 OD$_{8  }$ &  $ 9.52\pm0.20$  &  $ 0.108\pm0.020 $ \\
 (58475) & 1996 RE$_{11 }$ &  $12.22\pm0.10$  &  $ 0.085\pm0.031 $ \\
(192393)  & 1996 TT$_{22 }$ &  $ 7.80\pm0.06$  &  $ 0.088\pm0.009 $ \\
 (37789) & 1997 UL$_{16 }$ &  $14.22\pm0.22$  &  $ 0.062\pm0.009 $ \\
\nodata  & 1998 WM$_{24 }$ &  $ 6.90\pm0.16$  &  $ 0.094\pm0.009 $ \\
\nodata  & 1998 WO$_{39 }$ &  $10.20\pm0.12$  &  $ 0.104\pm0.010 $ \\
 (40262) & 1999 CF$_{156}$ &  $10.80\pm0.08$  &  $ 0.079\pm0.007 $ \\
 (59355) & 1999 CL$_{153}$ &  $ 9.10\pm0.08$  &  $ 0.102\pm0.009 $ \\
 (60257) & 1999 WB$_{25 }$ &  $10.66\pm0.08$  &  $ 0.155\pm0.056 $ \\
 (60322) & 1999 XB$_{257}$ &  $24.02\pm0.14$  &  $ 0.067\pm0.005 $ \\
(192942)  & 2000 AB$_{219}$ &  $ 7.72\pm0.04$  &  $ 0.118\pm0.011 $ \\
 (60388) & 2000 AY$_{217}$ &  $11.12\pm0.08$  &  $ 0.043\pm0.004 $ \\
(162396) & 2000 CV$_{120}$ &  $13.90\pm0.08$  &  $ 0.058\pm0.005 $ \\
 (60421) & 2000 CZ$_{31 }$ &  $12.98\pm0.08$  &  $ 0.050\pm0.005 $ \\
 (62692) & 2000 TE$_{24 }$ &  $18.38\pm0.12$  &  $ 0.073\pm0.005 $ \\
 (68112) & 2000 YC$_{143}$ &  $10.38\pm0.08$  &  $ 0.180\pm0.024 $ \\
 (63193) & 2000 YY$_{118}$ &  $13.70\pm0.22$  &  $ 0.094\pm0.017 $ \\
 (63259) & 2001 BS$_{81 }$ &  $ 9.96\pm0.08$  &  $ 0.094\pm0.008 $ \\
 (88240) & 2001 CG$_{21 }$ &  $12.16\pm0.16$  &  $ 0.108\pm0.018 $ \\
 (63284) & 2001 DM$_{46 }$ &  $10.26\pm1.32$  &  $ 0.252\pm0.050 $ \\
 (63279) & 2001 DW$_{34 }$ &  $ 8.32\pm0.58$  &  $ 0.213\pm0.022 $ \\
 (28960) & 2001 DZ$_{81 }$ &  $14.52\pm0.18$  &  $ 0.123\pm0.021 $ \\
(109266) & 2001 QL$_{110}$ &  $ 9.22\pm0.06$  &  $ 0.079\pm0.004 $ \\
(156222) & 2001 UB$_{91 }$ &  $ 7.34\pm0.06$  &  $ 0.120\pm0.008 $ \\
(156250) & 2001 UM$_{198}$ &  $ 9.04\pm0.06$  &  $ 0.084\pm0.012 $ \\
 (64326) & 2001 UX$_{46 }$ &  $17.28\pm0.10$  &  $ 0.066\pm0.002 $ \\
(158333) & 2001 WW$_{25 }$ &  $ 8.56\pm0.08$  &  $ 0.205\pm0.040 $ \\
\nodata  & 2002 CG$_{205}$ &  $ 7.64\pm0.14$  &  $ 0.141\pm0.010 $ \\
 (43627) & 2002 CL$_{224}$ &  $ 8.58\pm0.08$  &  $ 0.205\pm0.016 $ \\
 (65179) & 2002 CN$_{224}$ &  $10.60\pm0.02$  &  $ 0.113\pm0.012 $ \\
(166115) & 2002 CO$_{208}$ &  $ 6.66\pm0.06$  &  $ 0.111\pm0.010 $ \\
\nodata  & 2002 CS$_{266}$ &  $ 5.64\pm0.14$  &  $ 0.260\pm0.037 $ \\
 (65174) & 2002 CW$_{207}$ &  $11.78\pm0.08$  &  $ 0.167\pm0.052 $ \\
 (65206) & 2002 DB$_{13 }$ &  $11.72\pm0.10$  &  $ 0.127\pm0.038 $ \\
 (89913) & 2002 EC$_{24 }$ &  $12.66\pm0.10$  &  $ 0.113\pm0.005 $ \\
 (65211) & 2002 EK$_{1  }$ &  $ 7.20\pm0.28$  &  $ 0.334\pm0.020 $ \\
(195258)  & 2002 EN$_{52 }$ &  $12.00\pm0.06$  &  $ 0.041\pm0.004 $ \\
 (65227) & 2002 ES$_{46 }$ &  $14.04\pm0.10$  &  $ 0.179\pm0.003 $ \\
 (65217) & 2002 EY$_{16 }$ &  $12.70\pm0.08$  &  $ 0.165\pm0.028 $ \\
 (65250) & 2002 FT$_{14 }$ &  $10.72\pm0.10$  &  $ 0.074\pm0.007 $ \\
(183358) & 2002 VM$_{131}$ &  $ 9.62\pm0.06$  &  $ 0.108\pm0.006 $ \\
 (58096) & Oineus          &  $14.40\pm0.50$  &  $ 0.088\pm0.013 $ \\
\enddata
\tablecomments{Here 
$D$ is the effective diameter and $p_R$ is the Cousins R-band
geometric albedo. For both quantities, the quoted
error is a 1-$\sigma$ formal error given the modeling assumptions and
the photometric uncertainties.}
\end{deluxetable}

\clearpage

\begin{deluxetable}{lccccll}
\tablecolumns{7}
\tablewidth{0pc}
\tablecaption{Ensemble R-Band Geometric Albedos}
\tablehead{
\colhead{Group} & \colhead{$N$} & \colhead{Average}  
	& \colhead{Median} & \colhead{Std. Dev.} 
	& \colhead{Source} & \colhead{Excluding}}
\startdata
``Large" & 31 & $0.045\pm0.001$ & $0.044\pm0.001$ & 0.008 & Paper I &  
	1 outlier \\
``Small" & 44 & $0.121\pm0.003$ & $0.105\pm0.004$ & 0.062 & this work & 
	none \\
``Small" & 32 & $0.137\pm0.004$ & $0.117\pm0.005$ & 0.065 & this work & 
	no visible data \\
\enddata
\tablecomments{Here 
``large" and ``small" in the ``Group" column refer to Trojans of 
diameter greater
than 57 km and Trojans of diameter less than 24 km, respectively; 
$N$ is the number of objects included in that row's calculations;
and ``Excluding" indicates
which objects are excluded from that row's calculations. Error bars
on the averages come from propagating errors of the albedos
themselves; error bars on the medians come from Monte Carlo
simulations.}
\end{deluxetable}


\end{document}